\documentclass[aps,prb,twocolumn,superscriptaddress, showpacs]{revtex4-1}
\usepackage[utf8]{inputenc}
\usepackage{amssymb}
\usepackage{amsmath}
\usepackage{amsfonts}
\usepackage{bbold}
\usepackage[usenames,dvipsnames]{xcolor}
\usepackage[pdftex]{graphicx}
\usepackage{bm}
\usepackage[pdftex]{hyperref}
{\end{pmatrix}\end{medsize}} 
\usepackage{nccmath} 
\usepackage{natbib} 
\usepackage{physics} 

\hypersetup{colorlinks=true,linkcolor=red, citecolor=blue, urlcolor=blue,colorlinks=true,bookmarks=true,breaklinks=true}

\usepackage{array}
\newcolumntype{L}[1]{>{\raggedright\let\newline\\\arraybackslash\hspace{0pt}}m{#1}}
\newcolumntype{C}[1]{>{\centering\let\newline\\\arraybackslash\hspace{0pt}}m{#1}}
\newcolumntype{R}[1]{>{\raggedleft\let\newline\\\arraybackslash\hspace{0pt}}m{#1}}

\begin{document}
\title{Diagrammatic study of optical excitations in correlated systems}
\author{Olivier \surname{Simard}}
\affiliation{Department of Physics, University of Fribourg, 1700 Fribourg, Switzerland}
\author{Shintaro \surname{Takayoshi}}
\affiliation{Department of Physics, Konan University, Kobe 658-8501, Japan}
\author{Philipp \surname{Werner}}
\affiliation{Department of Physics, University of Fribourg, 1700 Fribourg, Switzerland}
\date{\today}


\begin{abstract}
The optical conductivity contains relevant information on the properties of correlated electron systems. In infinite dimensions, where dynamical mean field theory becomes exact, vertex corrections can be neglected and the conductivity  computed from particle-hole bubbles. An interesting question concerns the nature and effect of the most relevant vertex corrections in finite-dimensional systems.
A recent numerical study showed that the dominant vertex correction near an ordering instability with wave vector $\pi$ comes from a vertical ladder, analogous to the Maki-Thompson diagram. Since the RPA version of this ladder diagram, dubbed $\pi$-ton, can be easily evaluated, this suggests a simple procedure for incorporating antiferromagnetic or charge density wave fluctuations into dynamical mean field estimates of the optical conductivity and related susceptibilities. We implement this procedure for the half-filled Hubbard model, considering the $\pi$-ton and a double-ladder extension of the  $\pi$-ton, and reveal the spectral signatures of these vertex corrections.      
\end{abstract}
\maketitle


\section{Introduction}
\label{sec:Introduction}

Optical probes play an important role in the study of strongly correlated electron systems. In particular, the frequency-dependent optical conductivity contains a wealth of information on the correlated state, such as the energy gaps, carrier number, and kinetic energy. Different types of scattering processes and the formation of composite particles, such as excitons, also leave a trace in the optical conductivity. As a result, this observable has been extensively used to investigate correlated materials. It has played an important role especially in the study of the unconventional normal state of high-T$_c$ cuprates.\cite{Uchida_1991,Basov_2005}

In a diagrammatic language, the optical conductivity can be expressed as a particle-hole bubble plus vertex corrections. A recent numerical investigation by Kauch {\it et al.}\cite{kauch_pitons_2019} revealed that in correlated systems with strong fluctuations at momentum ${\bf k}_\pi=(\pi,\pi, \ldots)$, such as systems on a hypercubic lattice in the vicinity of an antiferromagnetic (AFM) or charge density wave (CDW) instability, the dominant vertex correction comes from a vertical ladder with momentum exchange ${\bf k}-{\bf k}'\approx {\bf k}_\pi$. In particular, this vertex correction should be relevant in the simplest model for high-T$_c$ cuprates, the square lattice single band Hubbard model near half-filling, because of the strong AFM correlations. The corresponding diagram, dubbed $\pi$-ton in Ref.~\onlinecite{kauch_pitons_2019}, has been shown to result in a broadening of the Drude feature or a shift in the gap edge of the conductivity spectrum, but disentangling its contribution from other processes in numerical data is challenging.  

An RPA-type $\pi$-ton ladder with bare interactions can be easily evaluated, so that the calculation of the corresponding vertex correction provides a potentially simple way of incorporating relevant fluctuations into the bubble approximation for the optical conductivity and other susceptibilities. Motivated by the insights of Kauch {\it et al.},\cite{kauch_pitons_2019} and the lack of systematic data on the importance of the $\pi$-ton in the Hubbard model, we present here a dynamical mean field theory (DMFT)\cite{Georges_1996} based study in which the RPA-$\pi$-ton type vertex correction to the optical conductivity and to a related spin-spin correlation function is compared to the bubble contribution and to a double-ladder extension of the $\pi$-ton. We find that in the weakly correlated metallic regime, the RPA-$\pi$-ton leads to a broadening of the Drude peak in the conductivity, and a characteristic in-gap feature, while the double-ladder extension has little qualitative effect. At stronger interactions, but still in the metallic regime, the Drude peak is significantly suppressed, and the in-gap feature grows. In the strongly correlated (Mott insulating) regime, the strategy of adding RPA-$\pi$-ton vertex corrections breaks down since the RPA ladder no longer has a pole near the DMFT phase boundary, 
and hence is not particularly large in the parameter regions with the strongest AFM or CDW fluctuations. 

Even though our DMFT based method does not capture the physics specific to one-dimensional (1D) systems, we also compare the results to the ground state correlation functions obtained by the density matrix renormalization group (DMRG) method,\cite{White1992PRL,Schollwock2011AnnPhys} which do not show any obvious $\pi$-ton related features in the conductivity and spin susceptibility. One reason is the presence of a Mott gap for any $U>0$, which leads to prominent peaks in the optical conductivity associated with charge excitations, in the same energy region where the $\pi$-ton contribution may be expected. 

The paper is organized as follows: The Hubbard model and the methods used to solve the DMFT equations are presented in Sec.~\ref{sec:Models_and_methods}. In Sec.~\ref{sec:correlation_functions} and Appendix~\ref{sec:appendice:Derivation_vertex_corrections} we derive the formulae for the considered ladder-type vertex corrections. The diagrammatic results for the half-filled Hubbard model and the comparison to DMRG are presented in Sec.~\ref{sec:results}. The discussion and conclusions can be found in Secs.~\ref{sec:discussion} and \ref{sec:conclusion}.


\section{Model and method}
\label{sec:Models_and_methods}

\subsection{Hubbard model}
\label{subsec:Hubbard_model}

The single-band Hubbard model with Hamiltonian 
\begin{align}
\label{eq:Hubbard_model_intro}
\hat{\mathcal{H}}=&-\sum_{ij,\sigma}t_{ij}\left(\hat{c}_{i,\sigma}^{\dagger}\hat{c}_{j,\sigma}+\text{H.c.}\right)+U\sum_i\hat{n}_{i,\uparrow}\hat{n}_{i,\downarrow}\nonumber\\
&-\mu \sum_i (\hat{n}_{i,\uparrow} + \hat{n}_{i,\downarrow})
\end{align} 
captures key aspects of strongly correlated electron systems. Here, the $t_{ij}$ are the hopping amplitudes between sites $j$ and $i$, $\sigma \in \{\uparrow,\downarrow\}$ denotes the spin, $\hat{c}^{(\dagger)}_{i,\sigma}$ are annihilation (creation) operators for site $i$, while $\hat{n}_{i\sigma}=\hat{c}^{\dagger}_{i,\sigma}\hat{c}_{i,\sigma}$ is the number operator, $U$ is the local repulsion and $\mu$ the chemical potential. We will consider nearest-neighbor hoppings and use the hopping $t$ as the unit of energy. We also set $\hbar$, $k_B$, the electric charge $e$ and the lattice spacings $a$ equal to unity. A half-filled system is obtained for $\mu=U/2$. 

\subsection{DMFT}
\label{subsec:DMFT}

\subsubsection{General formalism}

In DMFT, the lattice model is mapped to a single-site impurity model with a self-consistently determined bath of noninteracting electrons.\cite{Georges_1996} In this approximation the hybridization function mimics the effect of electrons hopping to neighboring sites of the lattice and back. The impurity Hamiltonian $\hat{\mathcal{H}}^{\prime}$ including the hybridization to the non-interacting bath reads
\begin{align}
\label{eq:Introduction:Impurity_Hamiltonian_DMFT}
\hat{\mathcal{H}}^{\prime} =& \hat{\mathcal{H}}_{\text{loc}}+ 
\sum_{i\alpha,\sigma}\left(\theta_{\alpha,\sigma}\hat{c}_{\sigma}^{\dagger}\hat{b}_{\alpha}+\text{H.c.}\right)+\sum_{\alpha,\sigma}\epsilon_{\alpha,\sigma}\hat{b}_{\alpha,\sigma}^{\dagger}\hat{b}_{\alpha,\sigma},
\end{align} 
where $ \hat{\mathcal{H}}_{\text{loc}}$ is the same local term as in the lattice model, $\hat{c}^{(\dagger)}$ annihilates (creates) an electron on the impurity and $\hat{b}^{(\dagger)}$ annihilates (creates) an electron in the bath. The environment is coupled to the impurity via the hybridizations $\theta_{\alpha,\sigma}$ with $\alpha$ labeling the non-interacting energy levels $\epsilon_{\alpha,\sigma}$. 
The impurity Green's function will be computed using a generalization of the iterated perturbation theory~\citep{arsenault_benchmark_2012,kajueter_new_1996} (IPT) method, introduced in Secs.~\ref{subsubsec:IPT}, and the non-crossing approximation~\citep{PhysRevB.24.4420,RevModPhys.59.845} (NCA) impurity solver, whose results will only be outlined (not shown). These methods are complementary in the sense that IPT should give reliable results at weak $U$, while NCA is appropriate for the strongly correlated regime. 

To compute the susceptibilities and their corrections, we work on the imaginary-time axis using fermionic (bosonic) Matsubara frequencies $\omega_n = 2\pi(n+1)/\beta$ ($\nu_n = 2\pi n/\beta$), where $n\in \mathbb{Z}$ and $\beta$ is the inverse temperature. Real frequency information will be obtained by Maximum Entropy analytical continuation.~\cite{bryan_maximum_1990}

\subsubsection{Paramagnetic self-consistency}
\label{subsubsec:PM}

In DMFT, the lattice self energy is assumed to be local and approximated by an impurity self energy 
$\Sigma_\sigma$. This allows to map the lattice system (\ref{eq:Hubbard_model_intro}) in a selfconsistent way onto the impurity model (\ref{eq:Introduction:Impurity_Hamiltonian_DMFT}). As a self-consistency condition, we impose that the interacting impurity Green's function 
$\mathcal{G}_{\sigma}(i\omega_n)$ is identical to the local lattice Green's function. 
This self-consistency condition fixes the bath related parameters in the impurity Hamiltonian, or (in an action formulation) the hybridization function $\Delta_\sigma$ resulting from Eq.~\eqref{eq:Introduction:Impurity_Hamiltonian_DMFT} when one integrates out the non-interacting bath electrons. This hybridization function plays the role of a dynamical mean field. Alternatively, one can define a so-called Weiss Green's function $\mathcal{G}_{\sigma}^{0}$, which is related to the hybridization function by 
\begin{align}
\label{eq:Models_and_methods:Weiss_Green}
\mathcal{G}_{\sigma}^{0}(i\omega_n) = \frac{1}{i\omega_n+\mu-\Delta_{\sigma}(i\omega_n)},
\end{align} 
and allows to express the impurity Green's function $\mathcal{G}$ by the impurity Dyson equation $\mathcal G_\sigma^{-1}(i\omega_n)=(\mathcal{G}_\sigma^0)^{-1}(i\omega_n)-\Sigma_\sigma(i\omega_n)$. $\Sigma[\mathcal{G}^0]$ will be computed using the impurity solver described in Sec.~\ref{subsubsec:IPT}.

Written in terms of the dynamical mean field $\Delta$, the DMFT self-consistency condition reads
\begin{align}
\label{eq:Models_and_methods:projected_green_function_impurity}
\mathcal{G}_{\sigma}[\Delta](i\omega_n) = \frac{1}{N}\sum_{\mathbf{k}}\frac{1}{i\omega_n + \mu - \epsilon(\mathbf{k}) - \Sigma_{\sigma}[\Delta](i\omega_n)},
\end{align}
where $\epsilon(\mathbf{k})=-2t\sum_{i=1}^D\cos{k_i}$ is the bare electronic dispersion on the hypercubic lattice in $D$ dimensions, and the self-energy is expressed in terms of $\mathcal{G}$ and $\Delta$ as
\begin{align}
\label{eq:Models_and_methods:impurity_self_energy}
\Sigma_{\sigma}[\Delta](i\omega_n) = i\omega_n + \mu - \mathcal{G}_{\sigma}[\Delta](i\omega_n)^{-1} - \Delta_{\sigma}(i\omega_n).
\end{align} Equations~\eqref{eq:Models_and_methods:projected_green_function_impurity} and \eqref{eq:Models_and_methods:impurity_self_energy} form a closed set of equations which determines $\Delta_\sigma$, and can be solved by iteration.  
For the paramagnetic solution, we impose $\Sigma_\uparrow=\Sigma_\downarrow$ and similarly for $\mathcal{G}$ and $\Delta$.

\subsubsection{Antiferromagnetic self-consistency}
\label{subsubsec:AFM}

DMFT can also treat two-sublattice type order on a bipartite lattice, such as antiferromagnetism.\cite{Georges_1996} 
We still assume local self-energies, but they may now be different on the two sublattices, and in the case of AFM order, they become spin dependent.  
Using Dyson's equation, and denoting the sublattices degrees of freedom by the indices $\{A,B\}$, $\mathcal{G}_{\sigma}^{\alpha\beta}(i\omega_n)^{-1}\mathcal{G}_{\sigma}^{\beta\alpha^{\prime}}(i\omega_n) = \delta_{\alpha,\alpha^{\prime}}$ becomes
\begin{align}
\label{eq:Models_and_methods:Dyson_equation_AFM}
&\begin{pmatrix}
i\omega_n + \mu - h\sigma - \Sigma^A_{\sigma}(i\omega_n) && -\epsilon(\mathbf{k})\\
-\epsilon(\mathbf{k}) && i\omega_n + \mu + h\sigma - \Sigma^B_{\sigma}(i\omega_n)
\end{pmatrix}\nonumber\\
&\times \begin{pmatrix}
\mathcal{G}_{\sigma}^{AA}(i\omega_n) && \mathcal{G}_{\sigma}^{AB}(i\omega_n) \\
\mathcal{G}_{\sigma}^{BA}(i\omega_n) && \mathcal{G}_{\sigma}^{BB}(i\omega_n)
\end{pmatrix} = \mathbb{1},
\end{align}
where $\epsilon(\mathbf{k})$ is the electronic dispersion restricted to the reduced Brillouin zone (rBZ), 
$h$ a constant staggered magnetic field that may be used as an initial perturbation in the DMFT self-consistency loop, and the $\Sigma^\alpha_{\sigma}$ are the sublattice local self-energies for spin $\sigma$. Hence, the local Green's functions for one sublattice impurity depend on the self-energy of the opposite sublattice as 
\begin{align}
\label{eq:Models_and_methods:G_bipartite_lattice_inversed}
&\mathcal{G}_{\sigma}^{\alpha\alpha}(i\omega_n) =\notag\\
&\sum_{{\bf k}\in \text{rBZ}} \left[i\omega_n+\mu-\Sigma_{\sigma}^{\alpha}(i\omega_n)-\frac{\epsilon(\mathbf{k})^2}{i\omega_n+\mu-\Sigma_{\sigma}^{-\alpha}(i\omega_n)}\right]^{-1},
\end{align} 
which follows from the inversion of 
Eq.~\eqref{eq:Models_and_methods:Dyson_equation_AFM}. Furthermore, since in N\'eel type AFM systems quantities such as $\mathcal{G}$ and $\Sigma$ obey the symmetry $\mathcal{X}^{-\alpha-\alpha}_{\sigma}=\mathcal{X}^{\alpha\alpha}_{-\sigma}$, the DMFT equations can be reduced to a single sublattice site with the two spin projections. Equation~\eqref{eq:Models_and_methods:G_bipartite_lattice_inversed} hence breaks down into a set of two coupled DMFT equations, one for each spin projection.

\subsubsection{IPT solver}
\label{subsubsec:IPT} 
Iterated perturbation theory (IPT) is a bare second-order perturbation theory for the Anderson impurity model~\eqref{eq:Introduction:Impurity_Hamiltonian_DMFT}.\cite{kajueter_new_1996,arsenault_benchmark_2012,tsuji_nonequilibrium_2013} It is exact in both the non-interacting and atomic limits, and provides the correct high-frequency behavior. The self-energy is approximated as 

\begin{align}
\label{eq:Models_and_methods:impurity_self_energy_IPT}
\Sigma_{\sigma}(i\omega_n) = U^2\int_0^{\beta}\mathrm{d}\tau e^{i\omega_n\tau}\mathcal{G}_{\sigma}^{0}(\tau)\mathcal{G}_{-\sigma}^{0}(\tau)\mathcal{G}_{-\sigma}^{0}(-\tau),
\end{align} where the Weiss Green's function is given by Eq.~\eqref{eq:Models_and_methods:Weiss_Green}. 
In addition, there may be a Hartree term, but in a half-filled paramagnetic state, this term can be absorbed into the chemical potential by choosing $\mu=U/2$. 

We can directly insert the self-energy \eqref{eq:Models_and_methods:impurity_self_energy_IPT} into Eq.~\eqref{eq:Models_and_methods:projected_green_function_impurity} and the impurity Dyson equation and iterate the solution until convergence. To interpolate the self-energy when Fourier transforming to fermionic Matsubara frequencies, a cubic spline is used.\footnote{See Appendix B of Ref.~\onlinecite{arsenault_benchmark_2012} and references therein for more details about the implementation.}

To break the spin rotation symmetry, one has to add to the self-energy~\eqref{eq:Models_and_methods:impurity_self_energy_IPT} the second-order Hartree term~\citep{tsuji_nonequilibrium_2013}

\begin{align}
\label{eq:Models_and_methods:impurity_self_energy_IPT_AFM_additional_term}
\Sigma^{(2H)}_{\sigma}(i\omega_n) = U^2n^0_{\sigma}\int_0^{\beta}\mathrm{d}\tau^{\prime}\mathcal{G}^0_{-\sigma}(\tau-\tau^{\prime})\mathcal{G}^0_{-\sigma}(\tau^{\prime}-\tau),
\end{align} where $n^0_{\sigma}=\mathcal{G}^0_{\sigma}(0^-)$. The latter term, which produces a relative shift of the chemical potential on the two sublattices, is important to obtain converged IPT solutions in the AFM state.


\subsection{Correlation functions}
\label{sec:correlation_functions}

In this section we explain the general formalism for computing the optical conductivity and related susceptibilities. We make use of Hedin's~\citep{Tremblay:notes,stefanucci_van_leeuwen_2013} equations to derive the $\pi$-ton ladder-type vertex corrections to the current-current and spin-spin correlation functions. 

The vertex corrections to various response functions can be computed using the Schwinger formalism.\citep{Tremblay:notes} Specifically, we are interested in charge and spin response functions for model \eqref{eq:Hubbard_model_intro} with local density-density interactions. For the sake of an efficient notation, we introduce numbers encapsulating space-time variables, i.e $1\equiv(\mathbf{x}_1,\tau_1)$, and use bars over the numbers to indicate a space-time integration:
\begin{align}
\label{eq:meaning_of_bars_over_numbers_h}
A(\bar{1})\equiv \int_0^{\beta}\mathrm{d}\tau_1\idotsint_{-\infty}^{\infty}\mathrm{d}^Dx_1 \ A(\mathbf{x}_1,\tau_1),
\end{align} 
with $D$ the spatial dimension(s) of the system. Greek letters represent the discrete electronic degrees of freedom such as spin and orbitals. Repeated Greek letters are implicitly summed over.

\begin{figure}[t]
  \centering
    \includegraphics[width=\linewidth]{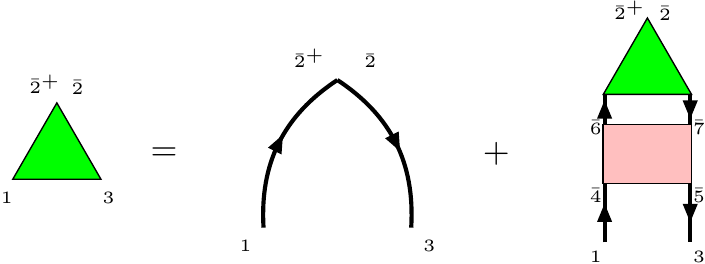}
      \caption{Diagrammatic representation of $\frac{\delta\mathcal{G}}{\delta\phi}$. The pink box represents $\frac{\delta\Sigma}{\delta\mathcal{G}}$. Note that for the Hubbard model and a pink box representing a vertical ladder, $\bar{4}=\bar{6}$ and $\bar{5}=\bar{7}$.}
  \label{fig:four_point_delta_G_delta_phi_diagrams}
\end{figure}

The functional $\mathcal{Z}$ that generates correlation functions is a modified partition function:
\begin{align}
\label{eq:correlation_functions:generating_functional}
\mathcal{Z}[\phi] = \text{Tr}\left[e^{-\beta\hat{K}}\mathcal{T}_{\tau}e^{-\hat{c}_{\alpha^{\prime}}^{\dagger}(\bar{1})\phi_{\alpha^{\prime}\beta^{\prime}}(\bar{1},\bar{2})\hat{c}_{\beta^{\prime}}(\bar{2})}\right],
\end{align} 
where $\phi$ is a source field, whose value has to be set to zero when computing physical quantities. The system is connected to both temperature and particle baths, so the grand-canonical ensemble is used and $\hat{K} = \hat{\mathcal{H}}-\mu\hat{N}$, with $\mu$ the chemical potential and $N$ the total number of particles. The corresponding imaginary-time Green's function is 
\begin{align}
\label{eq:correlation_functions:imaginary_time_Greens_function}
-\frac{\delta\ln{\mathcal{Z}[\phi]}}{\delta\phi_{\alpha\beta}(2,1)} = -\langle\mathcal{T}_{\tau}\hat{c}_{\alpha}(1)\hat{c}_{\beta}^{\dagger}(2)\rangle_{\phi} = \mathcal{G}^{\phi}_{\alpha\beta}(1,2),
\end{align} 
where the average value means
\begin{align}
\label{eq:correlation_functions:generating_functional}
\langle \cdots \rangle_{\phi} \equiv \text{Tr}\left[\frac{e^{-\beta\hat{K}}}{\mathcal{Z}[\phi]}e^{-\hat{c}_{\alpha^{\prime}}^{\dagger}(\bar{1})\phi_{\alpha^{\prime}\beta^{\prime}}(\bar{1},\bar{2})\hat{c}_{\beta^{\prime}}(\bar{2})}\cdots\right].
\end{align} 
By taking one more derivative with respect to the source field one can generate the four-point correlation function linked to the self-energy via the equations of motion and Dyson's equation
\begin{align}
\label{eq:correlation_functions:four_point_correlation_function}
\frac{\delta\mathcal{G}^{\phi}_{\alpha\beta}(1,3)}{\delta\phi_{\gamma\delta}(2^{+},2)} =& \mathcal{G}^{\phi}_{\delta\gamma}(2,2^{+})\mathcal{G}^{\phi}_{\alpha\beta}(1,3)\notag\\
&+\langle\mathcal{T}_{\tau}\hat{c}_{\alpha}(1)\hat{c}^{\dagger}_{\beta}(3)\hat{c}_{\gamma}^{\dagger}(2^{+})\hat{c}_{\delta}(2)\rangle_{\phi}.
\end{align} 
Another important ingredient is the identity relation which ensures that the generated Feynman diagrams of the self-energy are irreducible, coming from the fact that $\frac{\delta\left(\mathcal{G}^{\phi}_{\eta\beta}(1,\bar{4})\mathcal{G}_{\beta\theta}^{\phi}(\bar{4},3)^{-1}\right)}{\delta\phi_{\gamma\delta}(2^+,2)} = 0$:
\begin{align}
\label{eq:correlation_functions:identity_relation}
\frac{\delta\mathcal{G}^{\phi}_{\eta\Omega}(1,3)}{\delta\phi_{\gamma\delta}(2^+,2)} =& \mathcal{G}^{\phi}_{\eta\gamma}(1,2^+)\mathcal{G}^{\phi}_{\delta\Omega}(2,3)\delta_{\eta\gamma}\delta_{\Omega,\delta} \notag\\
&+\mathcal{G}^{\phi}_{\eta\beta}(1,\bar{4})\frac{\delta\Sigma^{\phi}_{\beta\theta}(\bar{4},\bar{5})}{\delta\mathcal{G}^{\phi}_{\Lambda\Gamma}(\bar{6},\bar{7})}\frac{\delta\mathcal{G}^{\phi}_{\Lambda\Gamma}(\bar{6},\bar{7})}{\delta\phi_{\gamma\delta}(2^+,2)}\mathcal{G}^{\phi}_{\theta\Omega}(\bar{5},3).
\end{align} 
Equation~\eqref{eq:correlation_functions:identity_relation} can be represented in terms of Feynman diagrams as shown in  Fig.~\ref{fig:four_point_delta_G_delta_phi_diagrams}. For the sake of conciseness and clarity the following notations will be employed on several occasions: $\frac{\delta\Sigma^{\phi}_{\sigma}(4,5)}{\delta\mathcal{G}^{\phi}_{\sigma^{\prime\prime}}(6,7)}\to \square^{\phi}_{\sigma\sigma^{\prime\prime}}\left(\frac{4,5}{6,7}\right)$ and $\frac{\delta\mathcal{G}^{\phi}_{\sigma^{\prime\prime}}(6,7)}{\delta\phi_{\sigma^{\prime}}(2^+,2)}\to \ \blacktriangleright^{\phi}_{\sigma^{\prime\prime}\sigma^{\prime}}(6,7,2)$. In the case of the Hubbard model, the first expression simplifies to 
$\square^{\phi}_{\sigma\sigma^{\prime\prime}}(4-5)\delta(4-6)\delta(5-7)\delta_{\sigma^{\prime\prime},-\sigma}$. Also, later on, to distinguish ladder-type vertex corrections whose right extremity terminates with a spin flip from those that don't, we will split up $\blacktriangleright$ into an even contribution containing an even number of vertical ladders (pink boxes) terminating with the same spin, denoted $\blacktriangleright^{(\text{even})}$ (see Fig.~\ref{fig:delta_G_delta_phi_diagrams_even}), and odd contribution containing an odd number of vertical ladders $\blacktriangleright^{(\text{odd})}$ terminating with a spin flip (see Fig.~\ref{fig:delta_G_delta_phi_diagrams_odd}). Both $\blacktriangleright^{(\text{even})}$ and $\blacktriangleright^{(\text{odd})}$ are discussed further in Appendix~\ref{sec:appendice:Derivation_vertex_corrections}.

We now combine Eqs.~\eqref{eq:correlation_functions:four_point_correlation_function} and \eqref{eq:correlation_functions:identity_relation} to derive the expressions for the density-density response, and based on this, the current-current response along direction $i\in\{x,y\}$ $\chi_{j_ij_i}$ and the spin-spin response $\chi_{S_zS_z}$. 
As detailed in Appendix~\ref{sec:appendice:Derivation_vertex_corrections}, the equations of motion provide an expression for $\Sigma\mathcal{G}$ (Eq.~(\ref{eq:appendice:Derivation_vertex_corrections:eq_motion_Hubbard_model})),
\begin{align}
\label{eq:appendice:Derivation_vertex_corrections:eq_motion_Hubbard_model_maintext}
&\Sigma_{\sigma}^{\phi}(1,\bar{2})\mathcal{G}_{\sigma}^{\phi}(\bar{2},2)=\notag\\
&\quad -\sum_{\sigma_2}U\delta_{\sigma_2-\sigma}\delta(1-\bar{2})\left<\mathcal{T}_{\tau}\hat{c}_{\sigma}(1)\hat{c}^{\dagger}_{\sigma}(2)\hat{c}^{\dagger}_{\sigma_2}(\bar{2}^+)\hat{c}_{\sigma_2}(\bar{2})\right>_{\phi}, 
\end{align}
which involves the same four-point correlation function as Eq.~\eqref{eq:correlation_functions:four_point_correlation_function}. We can combine Eqs.~\eqref{eq:correlation_functions:four_point_correlation_function}, \eqref{eq:correlation_functions:identity_relation} and \eqref{eq:appendice:Derivation_vertex_corrections:eq_motion_Hubbard_model_maintext}
to express the self-energy of the Hubbard model\footnote{The Fock term of the self-energy disappears due to the Pauli exclusion principle.} as
\begin{align}
\label{eq:correlation_functions:self_energy}
\Sigma^{\phi}_{\sigma}(1,3)&=U\mathcal{G}^{\phi}_{-\sigma}(1,1^+)\delta(1-3)\notag\\
&\phantom{=}-U\sum_{\sigma^{\prime}}\mathcal{G}^{\phi}_{\sigma}(1,\bar{4})\frac{\delta\Sigma^{\phi}_{\sigma}(\bar{4},3)}{\delta\mathcal{G}^{\phi}_{\sigma^{\prime}}(\bar{5},\bar{6})}\frac{\delta\mathcal{G}^{\phi}_{\sigma^{\prime}}(\bar{5},\bar{6})}{\delta\phi_{-\sigma}(1^+,1)},
\end{align} where the Greek indices have been traded for the spin $\sigma\in\{\uparrow,\downarrow\}$. Eq.~\eqref{eq:correlation_functions:self_energy} is illustrated in Fig.~\ref{fig:self_energy_diagrams}. The self-energy~\eqref{eq:correlation_functions:self_energy} will be used in conjunction with Eqs.~\eqref{eq:correlation_functions:four_point_correlation_function} and \eqref{eq:correlation_functions:identity_relation} to compute the susceptibilities.

\begin{figure}[t!]
  \centering
    \includegraphics[width=\linewidth]{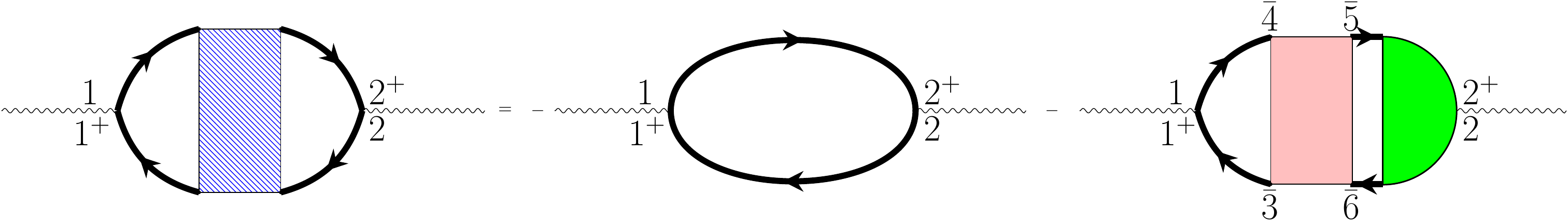}
      \caption{Diagrammatic representation of the dressed susceptibility. The pink box represents $\frac{\delta\Sigma}{\delta\mathcal{G}}$ and the green shape represents $\frac{\delta\mathcal{G}}{\delta\phi}$.
      }
  \label{fig:full_formula_for_susceptibility}
\end{figure}

From Eq.~\eqref{eq:correlation_functions:four_point_correlation_function}, we obtain the charge susceptibilities 
\begin{align}
\label{eq:correlation_functions:general_charge_susceptibility_expr}
&\chi^{\sigma\sigma^{\prime}}(1,1^+;2^+,2)=-\frac{\delta\mathcal{G}^{\phi}_{\sigma}(1,1^+)}{\delta\phi_{\sigma^{\prime}}(2^+,2)}\bigg\rvert_{\phi=0}\notag\\
&=\langle\mathcal{T}_{\tau}\hat{n}_{\sigma^{\prime}}(2)\hat{n}_{\sigma}(1)\rangle-\langle\hat{n}_{\sigma^{\prime}}(2)\rangle\langle\hat{n}_{\sigma}(1)\rangle\notag\\
&=\langle\mathcal{T}_{\tau}\left(\hat{n}_{\sigma^{\prime}}(2)-\langle\hat{n}_{\sigma^{\prime}}(2)\rangle\right)\left(\hat{n}_{\sigma}(1)-\langle\hat{n}_{\sigma}(1)\rangle\right)\rangle.
\end{align} 

Alternatively, one can express $\chi$ using Eq.~\eqref{eq:correlation_functions:identity_relation} as well as the Hubbard constraints as
\begin{align}
\label{eq:correlation_functions:general_charge_susceptibility_reexpr}
&\chi^{\sigma\sigma^{\prime}}(1,1^+;2^+,2)=-\mathcal{G}_{\sigma}(1,2^+)\mathcal{G}_{\sigma}(2,1^+)\delta_{\sigma,\sigma^{\prime}}\notag\\
&\hspace{5mm}-\sum_{\sigma^{\prime\prime}}\mathcal{G}^{\phi}_{\sigma}(1,\bar{4})\frac{\delta\Sigma^{\phi}_{\sigma}(\bar{4},\bar{3})}{\delta\mathcal{G}^{\phi}_{\sigma^{\prime\prime}}(\bar{5},\bar{6})}\frac{\delta\mathcal{G}^{\phi}_{\sigma^{\prime\prime}}(\bar{5},\bar{6})}{\delta\phi_{\sigma^{\prime}}(2^+,2)}\mathcal{G}^{\phi}_{\sigma}(\bar{3},1^+)\bigg\rvert_{\phi=0},
\end{align} whose second term, denoted $\chi_{\text{corr}}$, corresponds to vertex corrections and can be reexpressed with our notation as
\begin{align}
\label{eq:correlation_functions:general_charge_susceptibility_notation_corr}
&\chi_{\text{corr}}^{\sigma\sigma^{\prime}}(1,1^+;2^+,2) = \notag\\
& -\sum_{\sigma^{\prime\prime}}\mathcal{G}^{\phi}_{\sigma}(1,\bar{4})\mathcal{G}^{\phi}_{\sigma}(\bar{3},1^+)\square^{\phi}_{\sigma\sigma^{\prime\prime}}(\bar{4}-\bar{3})\blacktriangleright^{\phi}_{\sigma^{\prime\prime}\sigma^{\prime}}(\bar{4},\bar{3},2)\bigg\rvert_{\phi=0}.
\end{align}
A diagrammatic illustration of Eq.~\eqref{eq:correlation_functions:general_charge_susceptibility_reexpr} is given in Fig.~\ref{fig:full_formula_for_susceptibility}, and we will derive the formulae for the different susceptibilities $\chi_{j_ij_i}$ and $\chi_{S_zS_z}$ from this.

Because we are specifically interested in RPA-$\pi$-ton type vertex corrections, 
we consider two types of contributions, each of which involves a different collection of vertical ladder diagrams. We first treat the case involving single vertical ladder diagrams only, $\chi^{\sigma\sigma^{\prime}}_{\text{corr}}\to \chi^{\sigma\sigma^{\prime}}_{\text{sl}}$, for which a generic term is shown in Fig.~\ref{fig:single_ladder_diagrams_corr_sus}. The latter makes up the lowest order vertex correction comprising an odd number of vertical ladders, therefore leading to a spin flip once the photon is reemitted. We therefore use the first term of $\blacktriangleright^{(\text{even})}$ in Eq.~\eqref{eq:appendice:Derivation_vertex_corrections:infinite_ladder_full_dG_dphi_even}, since Eq.~\eqref{eq:correlation_functions:general_charge_susceptibility_notation_corr} already has one vertical ladder ($\square$). The formula is derived in Appendix~\ref{sec:appendice:Derivation_vertex_corrections}, so we only write out its final form here, using the 4-vector notation:

\begin{figure}[t]
  \centering
    \includegraphics[width=\linewidth]{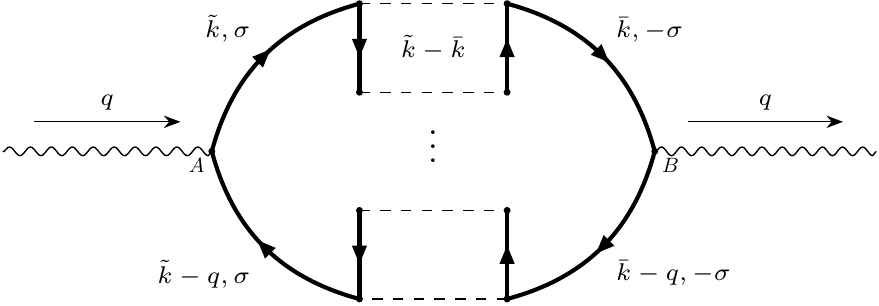}
      \caption{Illustration of the single-ladder vertex correction to the susceptibilities. All diagrams sharing this topology are summed up in Eq.~\eqref{eq:correlation_functions:single_ladder_k_space_representation_sus_corr}. To represent $\chi_{j_ij_i}$, the vertices $A$ and $B$ are both set equal to the velocity $v_i$, while for $\chi_{S_zS_z}$ they are set to the Pauli matrices $\sigma_z$.}
  \label{fig:single_ladder_diagrams_corr_sus}
\end{figure}

\begin{align}
\label{eq:correlation_functions:single_ladder_k_space_representation_sus_corr}
&\chi_{\text{sl}}^{\sigma-\sigma}(q) =\notag\\ 
&-\frac{U}{\left(\beta V\right)^2}\sum_{\tilde{k},\bar{k}}\frac{\mathcal{G}_{\sigma}(\tilde{k})\mathcal{G}_{\sigma}(\tilde{k}-q)\mathcal{G}_{-\sigma}\left(\bar{k}-q\right)\mathcal{G}_{-\sigma}(\bar{k})}{1+\underbrace{\frac{U}{\beta V}\sum_{\tilde{q}}\mathcal{G}_{\sigma}(\tilde{k}-\tilde{q})\mathcal{G}_{-\sigma}(\bar{k}-\tilde{q})}_{\equiv \chi^{\sigma-\sigma}_d(\tilde k-\bar k)}}\notag\\
&=-\frac{1}{(V\beta)^2}\sum_{\tilde{k},\bar{k}}\mathcal{G}_{\sigma}(\tilde{k})\mathcal{G}_{\sigma}(\tilde{k}-q)\square_{\sigma-\sigma}(\tilde{k}-\bar{k})\notag\\
&\hspace{23mm}\times\mathcal{G}_{-\sigma}(\bar{k})\mathcal{G}_{-\sigma}(\bar{k}-q).
\end{align} We denote the second term in the denominator by $\chi_d$ for latter purposes. In reciprocal space, after Fourier transforming Eq.~\eqref{eq:appendice:Derivation_vertex_corrections:functional_derivative_dself_energy_dG}, the $\square$ term reads 

\begin{align}
\label{eq:correlation_functions:single_ladder_k_space_square_expr}
\square_{\sigma-\sigma}(\tilde{k}-\bar{k}) = \frac{U}{1+\chi_d^{\sigma-\sigma}(\tilde{k}-\bar{k})}.
\end{align}

\begin{figure}[t]
  \centering
    \includegraphics[width=\linewidth]{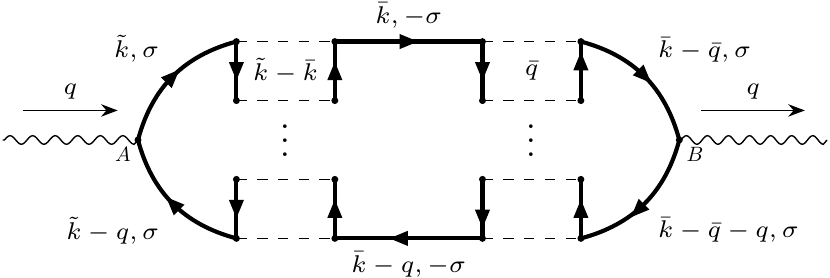}
      \caption{Illustration of the double-ladder vertex correction to the susceptibilities. Similarly to Fig.~\ref{fig:single_ladder_diagrams_corr_sus} the vertices $A$ and $B$ equal $v_i$ for $\chi_{j_ij_i}$, and $\sigma_z$ for $\chi_{S_zS_z}$.}
  \label{fig:infinite_ladder_diagrams_corr_sus}
\end{figure}

We will also consider the double-ladder case where the set of diagrams representing the vertex correction includes terms with two vertical ladders stacked together sideways ($\chi^{\sigma\sigma^{\prime}}_{\text{corr}}\to \chi^{\sigma\sigma^{\prime}}_{\text{dl}}$), as depicted in Fig.~\ref{fig:infinite_ladder_diagrams_corr_sus}. That set of diagrams sums up the lowest-order vertex corrections comprising an even number of vertical ladders: 
\begin{align}
&\chi_{\text{dl}}^{\sigma\sigma}(q)=\notag\\
&-\frac{1}{\left(\beta V\right)^3}\sum_{\substack{\tilde{k},\bar{k}\\ \bar{q}}}\mathcal{G}_{\sigma}(\tilde{k})\mathcal{G}_{\sigma}(\tilde{k}-q)\square_{\sigma-\sigma}(\tilde{k}-\bar{k})\mathcal{G}_{-\sigma}(\bar{k})\notag\\
&\hspace{5mm}\times\mathcal{G}_{-\sigma}(\bar{k}-q)\square_{-\sigma\sigma}(\bar{q})\mathcal{G}_{\sigma}(\bar{k}-\bar{q})\mathcal{G}_{\sigma}(\bar{k}-q-\bar{q}).
\label{eq:correlation_functions:infinite_ladder_k_space_representation_sus_corr}
\end{align}

In the case of the density-density correlation function, the vertices $A$ and $B$ in the diagrams \ref{fig:single_ladder_diagrams_corr_sus} and \ref{fig:infinite_ladder_diagrams_corr_sus} (as well as the particle-hole bubble) are identity operators, in the case of $\chi_{S_zS_z}$ they are spin operators $\frac{1}{2}\sigma_z$ and in the case of $\chi_{j_i j_i}$ the velocities $v_i(k)=\partial_{k_i}\epsilon(k)$ (since the charge is set to unity).  The longitudinal optical conductivity $\sigma_{ii}$ can be deduced from the imaginary component of the current-current correlation function (see Appendix~\ref{appendix:electronic_conductivity}) as

\begin{align}
\label{eq:correlation_functions:linear_response_longitudinal_conductivity}
\operatorname{Re}\sigma_{ii}(q_i,\omega) = \frac{\chi_{j_ij_i}^{\prime\prime}(q_i,\omega)}{\omega}.
\end{align}

\subsection{DMRG}

In order to provide accurate reference data for the 1D case, we also use the density matrix renormalization group (DMRG) method.\cite{White1992PRL,Schollwock2011AnnPhys} The calculation of the ground state is described in Appendix~\ref{appendix:DMRG}. Here, we briefly explain how to obtain the optical conductivity using the kernel polynomial method.\cite{Weisse2006RMP,Holzner2011PRB}

The optical conductivity is represented as 
$\sigma(\omega)=C(\omega)/\omega$, where 
\begin{align}
 C(\omega)
   =\langle \Psi_{0}|\hat{\mathcal{J}}\delta(\omega\hat{\mathbb{1}}-\hat{\mathcal{H}}+E_{0}\hat{\mathbb{1}})
     \hat{\mathcal{J}}|\Psi_{0}\rangle.
\end{align}
$\hat{\mathcal{J}}$ is the current operator, 
$|\Psi_{0}\rangle$ is the ground state and $E_{0}$ is its energy. 
We focus on the energy region $\omega \in [0,W]$, 
and map it to the interval 
$\omega^{\prime}=[-1+\epsilon_{\mathrm{s}},1-\epsilon_{\mathrm{s}}]$ 
($\epsilon_{\mathrm{s}}$ is a small safety factor, 
which is set to $0.0125$ in our study) 
through 
\begin{align}
 \omega^{\prime}=\frac{2(1-\epsilon_{\mathrm{s}})}{W}
   \omega-(1-\epsilon_{\mathrm{s}}).
\nonumber
\end{align}
The Hamiltonian is mapped to 
\begin{align}
 \hat{\mathcal{H}}^{\prime}=\frac{2(1-\epsilon_{\mathrm{s}})}{W}
   (\hat{\mathcal{H}}-E_{0}\hat{\mathbb{1}})-(1-\epsilon_{\mathrm{s}})\hat{\mathbb{1}}.
\nonumber
\end{align}
Since $-1<\omega^{\prime}<1$, the optical conductivity 
can be expanded as 
\begin{align}
 C(\omega')
   =&\langle \Psi_{0}|\hat{\mathcal{J}}
     \delta(\omega^{\prime}\hat{\mathbb{1}}-\hat{\mathcal{H}}^{\prime})
     \hat{\mathcal{J}}|\Psi_{0}\rangle\nonumber\\
   =&\frac{2(1-\epsilon_{\mathrm{s}})/W}{\pi\sqrt{1-{\omega^{\prime}}^{2}}}
   \Big[\mu_{0}+2\sum_{n=1}^{\infty}\mu_{n}T_{n}(\omega')\Big]
\label{eq:ChebyExpand}
\end{align}
by using the Chebyshev polynomials, 
\begin{align}
 T_{n}(\omega^{\prime})=\cos(n\arccos \omega^{\prime}).
\nonumber
\end{align}
The weight $\mu_{n}$ is calculated from 
$\bra{\Psi_{0}}\hat{\mathcal{J}}\ket{t_{n}}$, 
where 
$\ket{t_{n}}=T_{n}(\hat{\mathcal{H}}^{\prime})\hat{\mathcal{J}}\ket{\Psi_{0}}$. 
We evaluate $\mu_{n}$ using the matrix product state (MPS) method 
after obtaining the ground state by DMRG. 
For the evaluation, the recurrence relation 
of the Chebyshev polynomial is helpful, 
\begin{align}
 &|t_{0}\rangle=\hat{\mathcal{J}}|\Psi_{0}\rangle,\quad
 |t_{1}\rangle=\hat{\mathcal{H}}^{\prime}|t_{0}\rangle,\nonumber\\
 &|t_{n+2}\rangle
   =2\hat{\mathcal{H}}^{\prime}|t_{n+1}\rangle-|t_{n}\rangle.
\nonumber
\end{align}
In the numerical calculation, 
the expansion of Eq.~\eqref{eq:ChebyExpand} is carried out 
up to some fixed order $N$ and we multiply the Jackson damping factor~\cite{Weisse2006RMP} 
\begin{align}
 g_{n}=\frac{(N-n+1)\cos\frac{n\pi}{N+1}
   +\sin\frac{n\pi}{N+1}\cot\frac{\pi}{N+1}}{N+1}
\nonumber
\end{align}
to the weight $\mu_{n}$ as follows
\begin{align}
 C(\omega^{\prime})
   \simeq\frac{2(1-\epsilon_{\mathrm{s}})/W}{\pi\sqrt{1-{\omega^{\prime}}^{2}}}
   \Big[g_{0}\mu_{0}+2\sum_{n=1}^{N}g_{n}\mu_{n}T_{n}(\omega^{\prime})\Big].
\nonumber
\end{align}
In our study, the system size is 200 and 
the parameters are $W=15$ or 20, and $N=60$.


\section{Results}
\label{sec:results}

\subsection{General remarks}

We compute the longitudinal optical conductivity and the magnetic susceptibility for the weakly interacting half-filled one-band Hubbard model~Eq.~\eqref{eq:Hubbard_model_intro} in dimension $D=1$ using the DMFT Green's functions obtained with IPT~(Sec.~\ref{subsubsec:IPT}). 
DMFT produces results representative of high-dimensional systems, irrespective of $D$, so our choice of $D=1$ mainly serves to reduce the computational cost of the momentum summations. Qualitatively similar results have been obtained for $D=2$, but will not be explicitly discussed. Furthermore, the existence of AFM long-range order at $T_N>0$ in the DMFT solution is representative of $D\ge 3$ and we should thus regard the following diagrammatic results as characteristic properties of high-dimensional Hubbard models in the vicinity of the AFM phase boundary.  

We have also computed the $\pi$-ton-type vertex corrections in the strongly correlated (Mott) regime using NCA as impurity solver, but in this regime the RPA-type vertex correction is not meaningful, since it yields large values at high temperatures, far away from the AFM phase boundary. We thus restrict our attention to the weak-correlation regime, where the poles in the $\pi$-ton expressions can be shifted to the actual AFM boundary via modest corrections of the bare interaction. 

It is worth noting that since IPT and NCA are self-consistent methods that capture local correlations, the results should fulfill conservation laws. All the bare longitudinal optical conductivities presented in this paper obey the sum rule Eq.~\eqref{eq:appendix:electronic_conductivity:linear_response_current_density_key_rel_commutator_eval} within at least three digits.

\subsection{Phase diagram and renormalized couplings}
\label{subsec:phase_diagram_and_renorm_couplings}
 
\begin{figure}[b]
    \includegraphics[clip=true,trim=0cm 0cm 0cm 0cm, width=\columnwidth ]{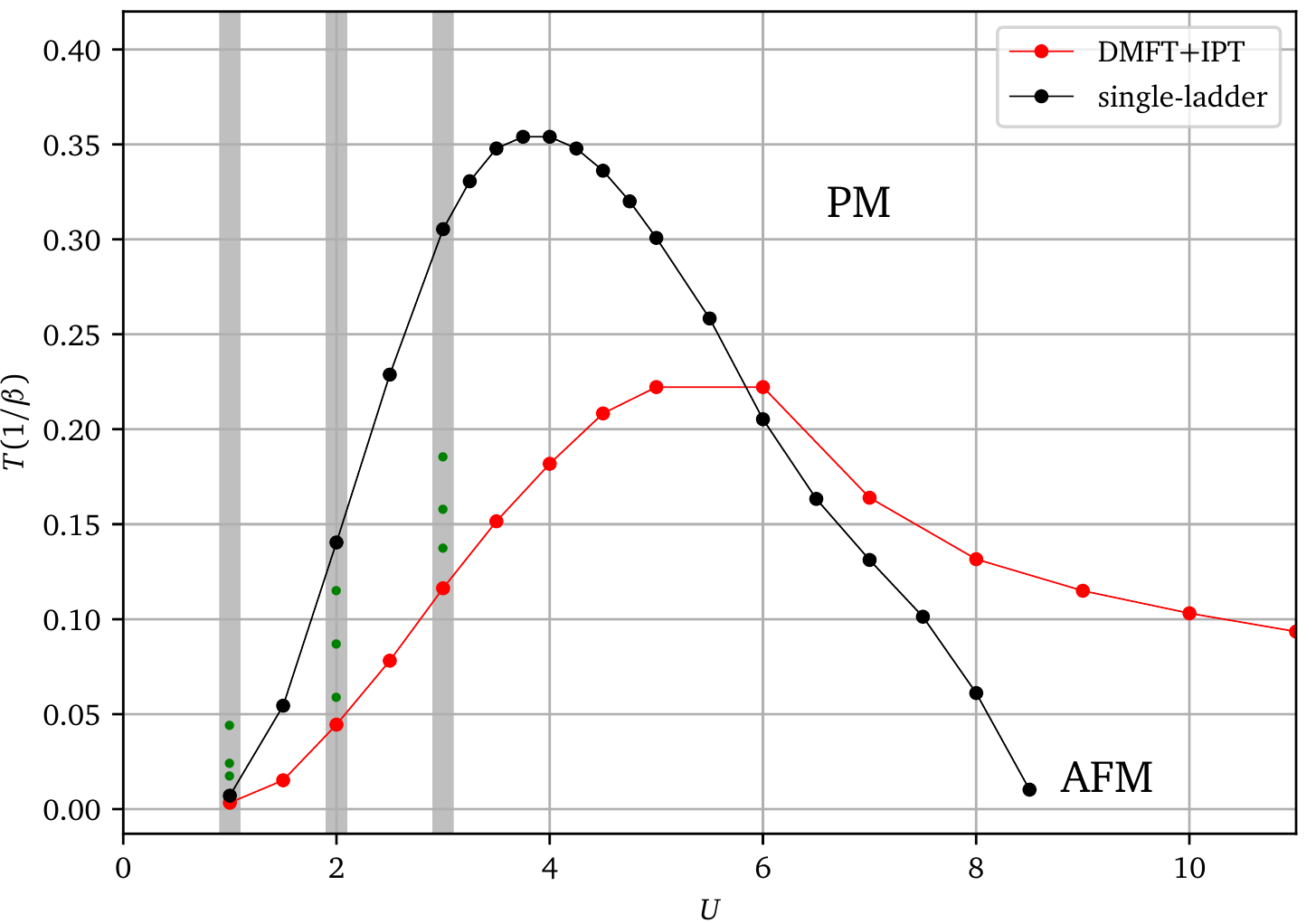}
    \caption{AFM phase boundary (red line) obtained with the IPT solver in the space of $U$ and $T$ at half-filling. The black line indicates the temperatures corresponding to the largest 
    single-ladder vertex corrections.  
    }
    \label{fig:results:1D:IPT:phase_diagram}
\end{figure}

To identify the parameter regions with strong AFM fluctuations, which are expected to enhance the $\pi$-ton type vertex corrections, we first map out the AFM phase boundary at half-filling. The DMFT phase diagram computed with the IPT solver is shown in Fig.~\ref{fig:results:1D:IPT:phase_diagram}. 

In order to calculate a lattice susceptibility that diverges at the DMFT phase boundary, one would have to compute a local vertex from the impurity model and use this as an approximation for the vertex of the lattice model in the solution of a Bethe-Salpeter equation.\cite{Jarrell_1992,Hoshino_2015} If we use the DMFT Green's functions in an RPA-type ladder, the corresponding susceptibility is not guaranteed to diverge, or become large, in the vicinity of the phase boundary. To figure out in which interaction regime the RPA-$\pi$-ton approach might produce meaningful results, we plot in Fig.~\ref{fig:results:1D:IPT:phase_diagram} the ($U$,$T$) parameters, where the denominator of the single-ladder diagram (Eq.~\eqref{eq:correlation_functions:single_ladder_k_space_representation_sus_corr}) at  $\tilde {\bf k}-\bar {\bf k}=\pi$ and $\omega_{n=0}$ vanishes (black line). 
Within DMFT+IPT, at small $U$, the black line remains close to the AFM phase boundary (it essentially follows the Hartree phase boundary\cite{tsuji_nonequilibrium_2013}). At $U\approx 4$ it reaches a maximum of $T\approx 0.35$, which is almost 50\% higher than the maximum N\'eel temperature ($T_N$), and then drops to small values faster than the AFM boundary. This drop, while qualitatively similar to the shape of the phase boundary, is not found in DMFT+NCA, which should provide a more accurate description on the ``Mott insulating" side of the AFM dome. In DMFT+NCA, the temperature associated with the dominant ladder contribution increases with increasing $U$. Hence, it is only meaningful to analyze the RPA-$\pi$-ton corrections to DMFT susceptibilities in the weak-coupling regime ($U\lesssim 3$), and we will thus from now on focus on the DMFT+IPT results. (We have checked our data against numerically exact DMFT results obtained with a continuous-time Monte Carlo method,\cite{Gull_2011} but found no qualitative differences to DMFT+IPT in this regime.)  

\begin{figure*}[ht]
    \includegraphics[clip=true,trim=0cm 0cm 0cm 0cm, width=1.3\columnwidth ]{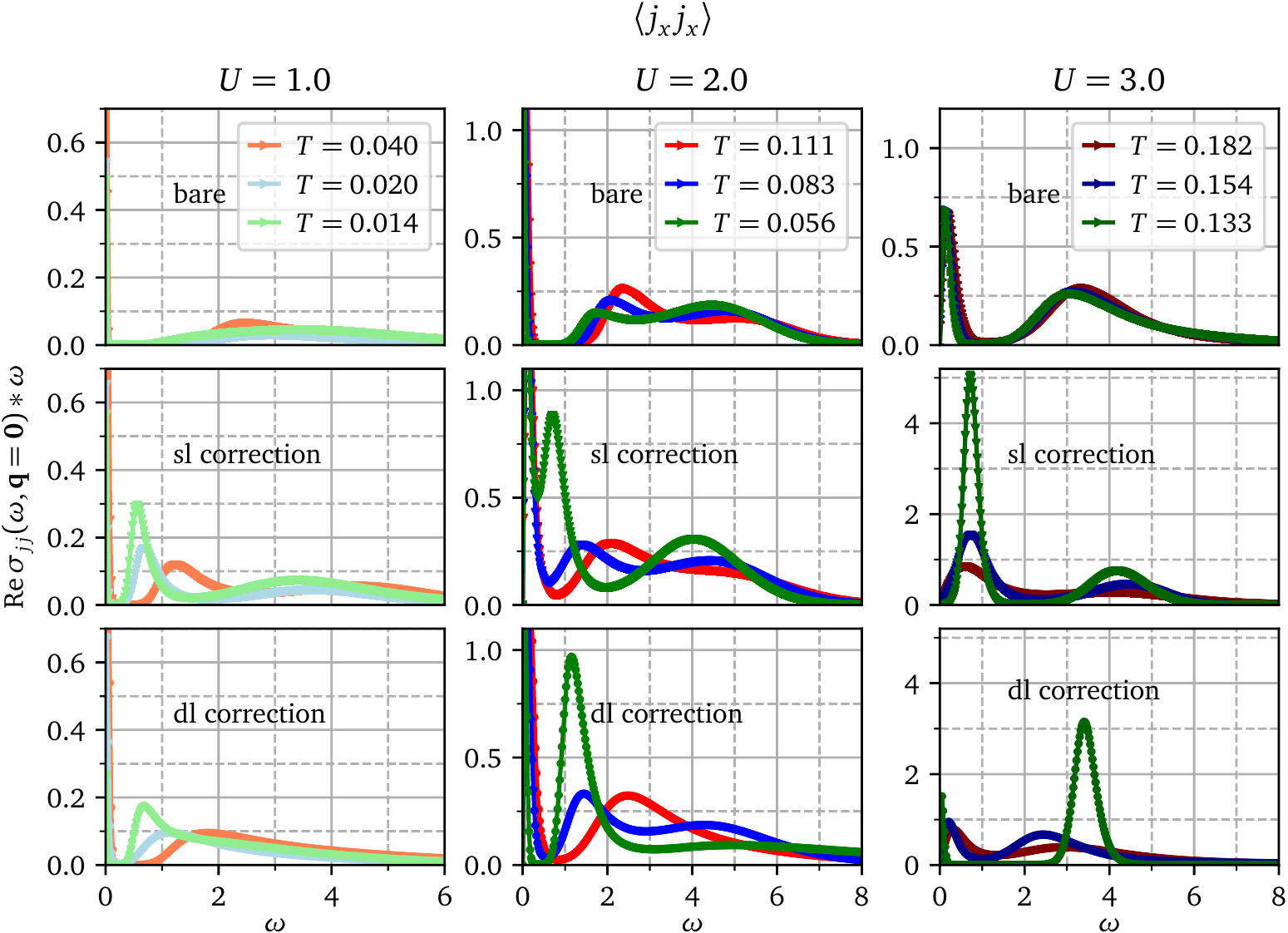}
    \caption{1D longitudinal optical conductivity for $U=1$ (first column), $U=2$ (second column) and $U=3$ (third column) obtained using DMFT+IPT and (for the ladder corrections) appropriately renormalized interactions. Top panel: bare response. Middle panel: bare response plus single-ladder vertex corrections. Bottom panel: bare response plus single-ladder and double-ladder vertex corrections. The temperatures considered for the different interactions are shown in the legends.
    }
    \label{fig:results:1D:IPT:jj}
\end{figure*}

Because the dominant vertex corrections from the single-ladder and double-ladder diagrams appear at temperatures which may be quite far from the N\'eel temperature, we introduce renormalized interactions $U^\text{ren}$. This renormalized coupling, which depends on $U$, ensures that the pole in the single-ladder expression 
is shifted to $T_N$ at the given $U$. (The DMFT Green's functions which enter the susceptibility calculations are computed with the unrenormalized $U$.) At $U=3$, we find $U^\text{ren}=U/1.4$, while at $U=2$, we have $U^\text{ren}=U/1.33$. At $U=1$, $U^\text{ren}$ is close enough to $T_N$ that we don't need to consider a renormalized coupling. 
\footnote{In the case of the double-ladder correction, even though the largest contribution may be shifted, we use the same $U^\text{ren}$, to enable a meaningful comparison.} 

These renormalizations of the bare interaction may be regarded as a consequence of the fact that the $\pi$-ton diagrams discussed in Ref.~\onlinecite{kauch_pitons_2019} are not RPA-type ladders, but involve a nontrivial vertex. The procedure is similar in spirit to the Kanamori theory for itinerant ferromagnetism, where a renormalized interaction is used in the mean-field Stoner condition.~\cite{10.1143/PTP.30.275}

\subsection{Optical conductivity and $q=0$ spin susceptibility}

In the following, we focus on $U=1$, $2$ and $3$ and compute both the optical conductivity and magnetic susceptibility for three temperatures approaching the phase boundary (see green dots in Fig.~\ref{fig:results:1D:IPT:phase_diagram}). The Green's functions entering the bubble and $\pi$-ton diagrams are DMFT Green's function for the corresponding $U$, while the interactions in the ladder expressions are renormalized as discussed in Sec.~\ref{subsec:phase_diagram_and_renorm_couplings}. The real-frequency spectra are computed using the Maximum Entropy method.\cite{bryan_maximum_1990}

\begin{figure*}[ht]
    \includegraphics[clip=true,trim=0cm 0cm 0cm 0cm, width=1.3\columnwidth ]{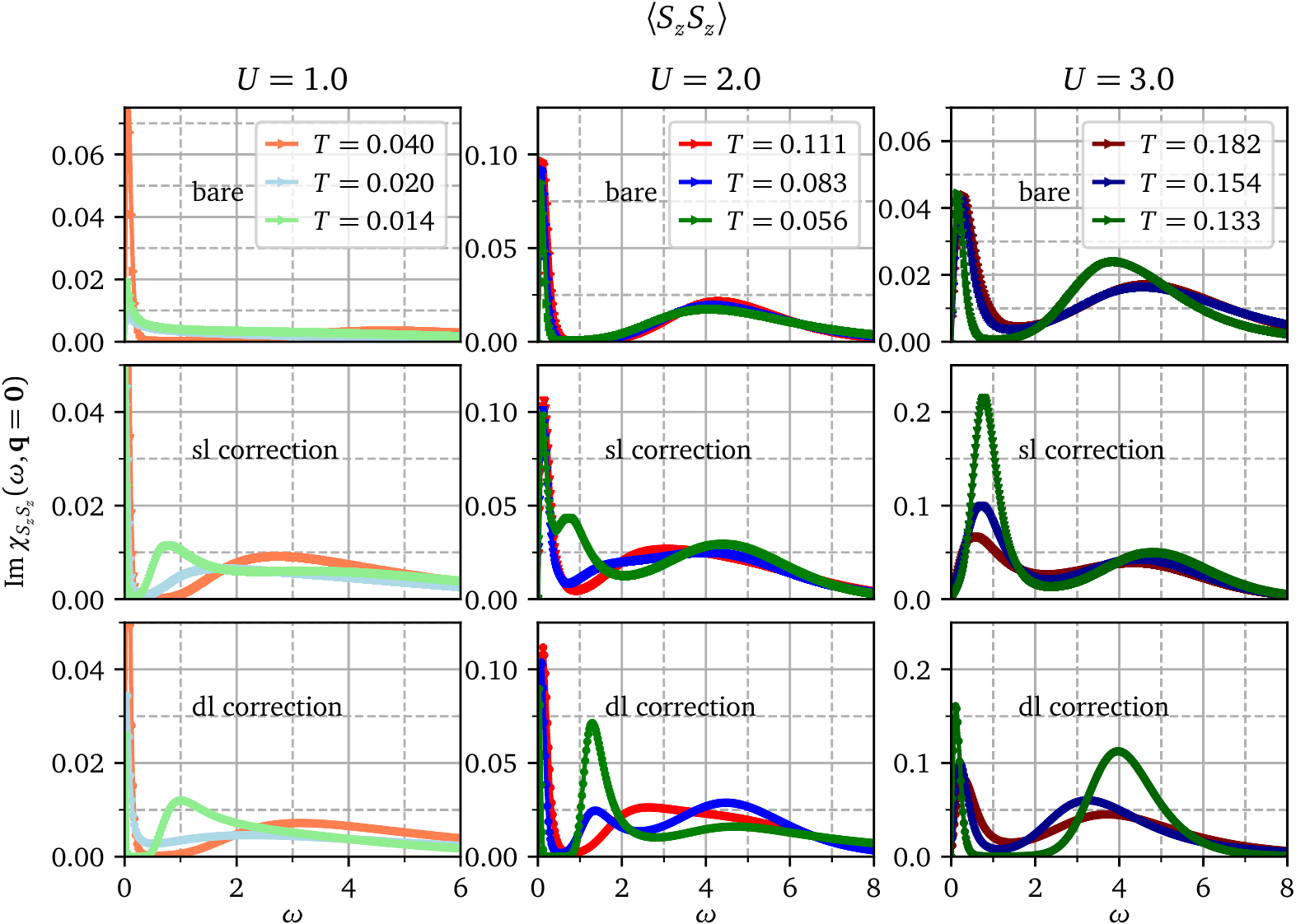}
    \caption{1D magnetic susceptibility for $U=1$ (first column), $U=2$ (second column) and $U=3$ (third column) obtained using DMFT+IPT and (for the ladder corrections) appropriately renormalized interactions. Top panel: bare response. Middle panel: bare response plus single-ladder vertex corrections. Bottom panel: bare response plus single-ladder and double-ladder vertex corrections. The temperatures considered for the different interactions are shown in the legends.
    }
    \label{fig:results:1D:IPT:szsz}
\end{figure*}

Figure~\ref{fig:results:1D:IPT:jj} shows the $\langle j_x j_x\rangle$ data for $U=1$ and $T=0.04,0.02,0.014$, $U=2$ and $T=0.111,0.083,0.056$, and $U=3$ and $T=0.182,0.154,0.133$, while Fig.~\ref{fig:results:1D:IPT:szsz} shows the $\langle S_z S_z\rangle$ data for the same parameter sets. In both figures, each column shows for the indicated value of $U$ starting from the top panel and going down: i) the bare response, ii) the bare response plus the single-ladder vertex corrections and iii) the bare response plus the single-ladder and double-ladder vertex corrections. The bubble contributions to the susceptibilities (top panels) exhibit a peak at small $\omega$ and a weak hump feature near $\omega\approx 4$, which originates from the peaks in the 1D density of states. For $U=2$ and $3$, there is also spectral weight around $\omega\approx U$, coming from Hubbard satellites in the density of states. As the temperature is lowered, the Drude peak of the optical conductivity becomes very narrow and sharp. In a Fermi liquid, $\sigma_\text{Drude}(\omega) \propto \gamma/[\pi(\gamma^2+\omega^2)]$ with $\gamma\sim T^2$ the scattering rate. Hence, at $U=1$ and low $T$, we have an almost $\delta$-function-like peak in the conductivity at $\omega=0$. To suppress this peak and highlight the structures at higher energies we plot $\text{Re} \sigma_{jj}(q=0,\omega)\ast\omega=\text{Im}\chi_{jj}(q=0,\omega)$. Also in the case of the spin susceptibility, we plot $\text{Im}\chi_{S_zS_z}(q=0,\omega)$. As $U$ increases, the scattering rate increases, the Drude peak becomes wider and the features associated with the  Hubbard subbands become more prominent. 

The effects of the single-ladder vertex corrections~(c.f.~Eq.~\eqref{eq:correlation_functions:single_ladder_k_space_representation_sus_corr}) on the optical conductivity are illustrated in the middle panels of Fig.~\ref{fig:results:1D:IPT:jj}. This vertex correction contributes a peak at $\omega\approx 0.6$, which grows as we approach $T_N$. Since this energy is larger than the width of the Drude peak, but smaller than the bandwidth  and $U$-related features, the $\pi$-ton appears as an in-gap peak in the optical conductivity. For small $U$, it also results in a broadening of the Drude peak, while at $U=3$, there are strong cancellations between the bare bubble and $\pi$-ton vertex correction, which suppress the Drude peak. At higher temperatures, the $\pi$-ton peak shifts to higher energies and merges with the high energy spectral weight of the bare bubble. Here, the effect of the vertex correction is a broadening of the Drude peak and a red-shift of the edge of the high-energy spectral weight. These results from our RPA-$\pi$-ton post-processing procedure look qualitatively consistent with the data presented in Ref.~\onlinecite{kauch_pitons_2019} which (for a set of different models) reported similar broadenings, in-gap peaks, and shifts of the gap edge.

The lowest panels illustrate the effect of the double-ladder vertex corrections~(c.f.~Eq.~\eqref{eq:correlation_functions:infinite_ladder_k_space_representation_sus_corr}) on the optical conductivity. For $U=1$ and $U=2$, including the double-ladder diagrams in the vertex corrections suppresses the hump at $\omega\simeq 4$ associated with the 1D density of states and broadens the peak associated with the in-gap state. 
We also find a shift of the $\pi$-ton peak to slightly higher energy. While the peak is broadened for $U=1$, it is not much affected in the case of $U=2$. Overall, the addition of a second vertical ladder has little qualitative effect on the $\pi$-ton for  $U\le 2$. For $U=3$ we find significant changes induced in the spectra as a result of the double-ladder corrections. The $\pi$-ton peak disappears and spectral weight is shifted to the high-energy and Drude features. We interpret this as a signature of a breakdown of the RPA-ladder post-processing approach, due to an increasing importance of various types of diagrams and the need for separate $U^\text{ren}$ for different types of corrections.

The results for the magnetic susceptibility, shown in Fig.~\ref{fig:results:1D:IPT:szsz}, are similar to those for the optical conductivity. We again see the broadening of the ``Drude peak" with increasing $U$ in the bare bubble contribution, and the appearance of high-energy spectral weight associated with Hubbard satellites. The single-ladder $\pi$-ton vertex correction yields an enhancement of the Drude feature at $U=1$, a broadening at $U=2$ and a suppression of the Drude peak at $U=3$, while characteristic in-gap peaks appear near $T_N$ around $\omega=0.8$. These $\pi$-ton peaks are slightly less prominent in the spin susceptibility than in the longitudinal optical conductivity.

The bottom panels illustrate the effect of the double-ladder vertex corrections for the various values of $U$. As in the case of the optical response, the additional set of diagrams does not significantly alter the main signature of the $\pi$-ton for $U=1$ and $U=2$. The amplitude of the peak is similar, while its position in the in-gap region is shifted slightly up. At $U=3$, the $\pi$-ton feature again disappears as a result of the double-ladder correction. Its spectral weight either merges with the hump produced by the 1D density of states or with the Drude feature. 

In connection with these spectra we should note that the positivity of the spectral weight is not a priori guaranteed. However, strong non-causal features should be detectable by Pad\'e analytical continuation,~\cite{vidberg_analytical_continuation_1977} and would be evident already on the Matsubara axis in the form of a non-monotonic $\omega_n$-dependence.\cite{Nilsson_2017} Since neither are observed for $U \le 3$, it is valid to use maximum entropy analytical continuation,\cite{bryan_maximum_1990} which enforces the positivity of the spectra.

\begin{figure}[ht]
    \includegraphics[clip=true,trim=0cm 0cm 0cm 0cm, width=\columnwidth ]{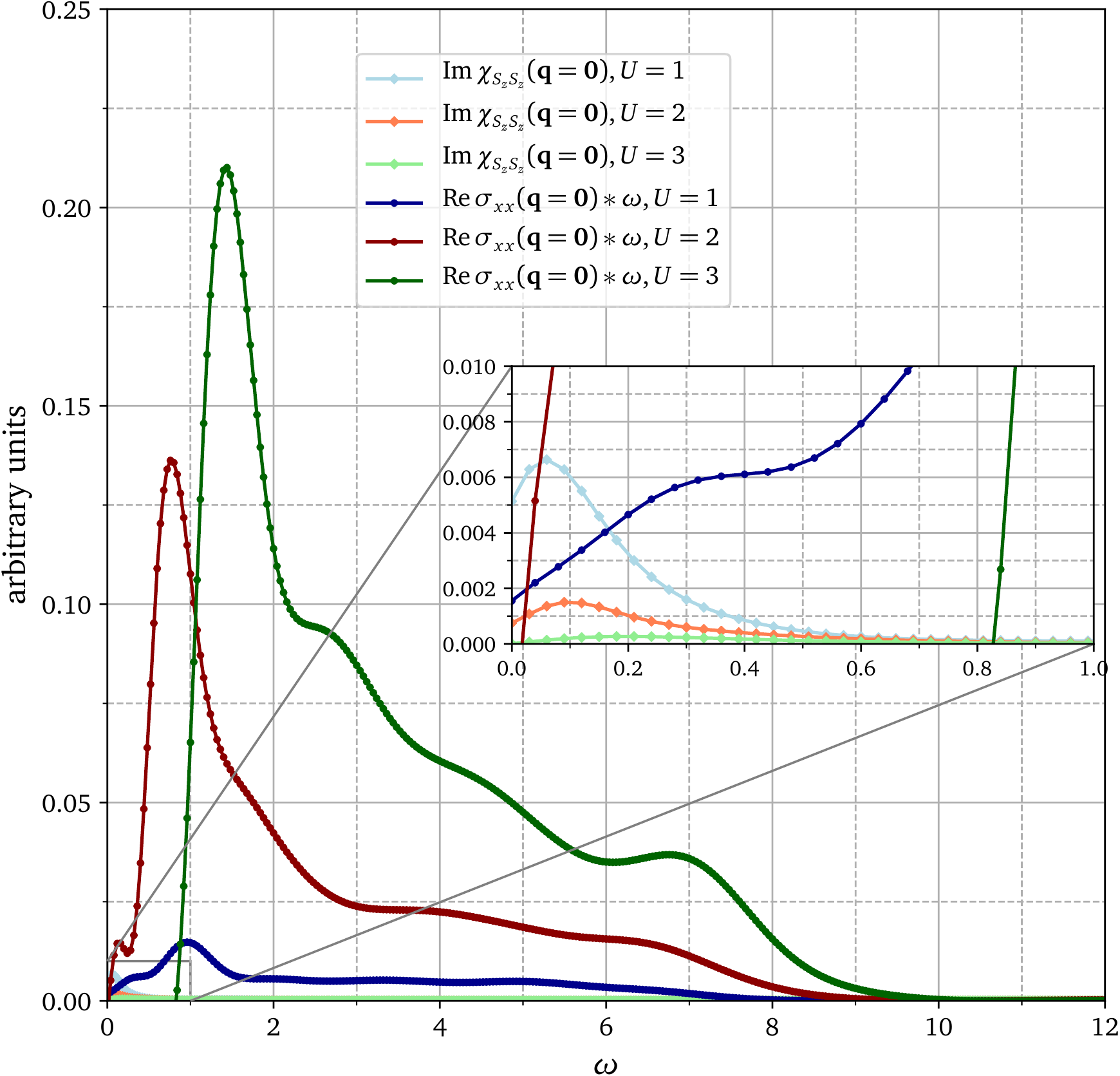}
    \caption{Longitudinal optical conductivity and magnetic susceptibility for $U=1,2$, $3$ and $T=0$ obtained using DMRG.}
    \label{fig:results:1D:DMRG:szsz_jj}
\end{figure}

\subsection{Comparison to DMRG}
\label{subsec:results:1D:DMRG}

We now compare the DMFT+IPT results of the previous section to the $T=0$ data obtained from DMRG for the 1D Hubbard model. DMRG is more accurate and captures nonlocal correlations, as well as 1D specific physics such as spin-charge separation.\cite{PhysRevB.90.155104} While we cannot expect a very close agreement between the DMFT susceptibilities, which are representative of finite-temperature higher dimensional systems, and the zero temperature DMRG results, 
it is nevertheless interesting to ask if the $\pi$-ton related features identified in the previous subsection leave some traces in the DMRG spectra.

We show in Fig.~\ref{fig:results:1D:DMRG:szsz_jj} the DMRG results for both the magnetic and current responses for the same values of the interaction $U$ as in the previous section. For the two susceptibilities, we attribute the spectral weight in the energy range $2\lesssim \omega\lesssim 7$ to structures in the 1D density of states, essentially captured at the level of the bare bubble in the diagrammatic calculation. Overall, the spin susceptibility in DMRG has lower spectral weight compared to the optical conductivity, consistent with the bare bubble calculations (compare Figs.~\ref{fig:results:1D:IPT:jj} and \ref{fig:results:1D:IPT:szsz}). The latter can be attributed to the factors $\tfrac12$ in the spin vertices, and the fact that the velocities at the vertices entering the optical conductivity --- corresponding to the derivative of the bare electronic dispersion with respect to momentum --- are proportional to a sinusoidal function weighted by $(2t)^2$, with maxima at $\mathbf{k}=\pm\frac{\pi}{2}$, which coincides with the momenta for which the spectral weight of the interacting Green's function is large, i.e where the self-energy only results in weak broadening (c.f.~Sec.~\ref{sec:discussion}).

The peak in $\text{Im}\chi_{S_zS_z}$ looks similar to the broadened ``Drude" feature found in the DMFT+IPT spectra with single-ladder and double-ladder corrections. However, the amplitude of the magnetic response is much weaker than that obtained in DMFT and the trend as a function of interaction is opposite: in DMFT the optical conductivity and magnetic susceptibility follow qualitatively similar trends, but in DMRG the spin-spin correlation function decreases with increasing $U$. 

In 1D, the low energy effective theory describes an independent sum of
electronic and spin degrees of freedom, which is known as the spin-charge separation.
For any $U>0$, the charge sector is in the Mott gapped phase,
and the spin sector is described by the Heisenberg model.
The exchange coupling is proportional to $t \sqrt{1- \mathrm{const} (U/t)}$
in the weak $U$ regime, which connects to $4t^2 /U$ in the strong
$U$ regime.
Hence, the spin exchange coupling decreases with increasing $U$.

A prominent peak appears at low frequencies ($\omega\approx 0.2 - 1.2$) in the optical conductivity. This peak moves up in frequency and increases in amplitude with increasing $U$, a behavior qualitatively similar to the $\pi$-ton peak identified in the diagrammatic analysis at approximately the same energies (Fig.~\ref{fig:results:1D:IPT:jj}). While one might thus expect a significant $\pi$-ton contribution, this peak in the DMRG solution is mainly originating from charge excitations across the Mott gap.~\cite{giamarchi2004quantum} One peculiarity of the half-filled 1D case is that it is Mott insulating at zero temperature for any $U>0$. (The absence of a gap in the $U=1$ and $2$ spectra is due to broadening.) Hence, even if a $\pi$-ton feature exists in the energy range suggested by the ladder calculations, it is dominated by the Mott gap feature in 1D. At $U=3$, our DMFT results for both the magnetic and optical responses show that the single-ladder vertex corrections almost completely suppress the Drude peak at $\omega=0$. (It is not completely suppressed when considering the double-ladder vertex corrections.) 
This suppression is not observed in DMRG in the case of the magnetic response. The qualitative difference between the spin and charge responses in DMRG may be attributed to specificities of the 1D and $T=0$ physics of the Hubbard model. In the 1D case, the magnetic excitations created by the $S_z$ operator are gapless because there is no AFM order even at $T=0$. 
In higher dimensions, the magnetic excitations by $S_z$ are gapped in the AFM phase, where the Drude peak disappears. The $\pi$-ton diagram captures the effect of strong AFM correlations, and thus leads to the suppression of the Drude peak. 

The spin and charge sectors of the 1D system become less asymmetric with increasing temperature,\cite{PhysRevB.90.155104} and we expect the spectrum for the spin correlation function to look more similar to that for the charge correlation function. Indeed, the optical conductivity in 1D features a Drude peak at elevated temperatures,\cite{PhysRevB.90.155104,PhysRevLett.65.243} and it would be interesting to perform a comparison between the diagrammatic results of the previous section and $T>0$ DMRG results, which is however beyond the scope of the present study. 


\begin{figure}[t]
    \includegraphics[clip=true,trim=0cm 0cm 0cm 0cm, width=\columnwidth ]{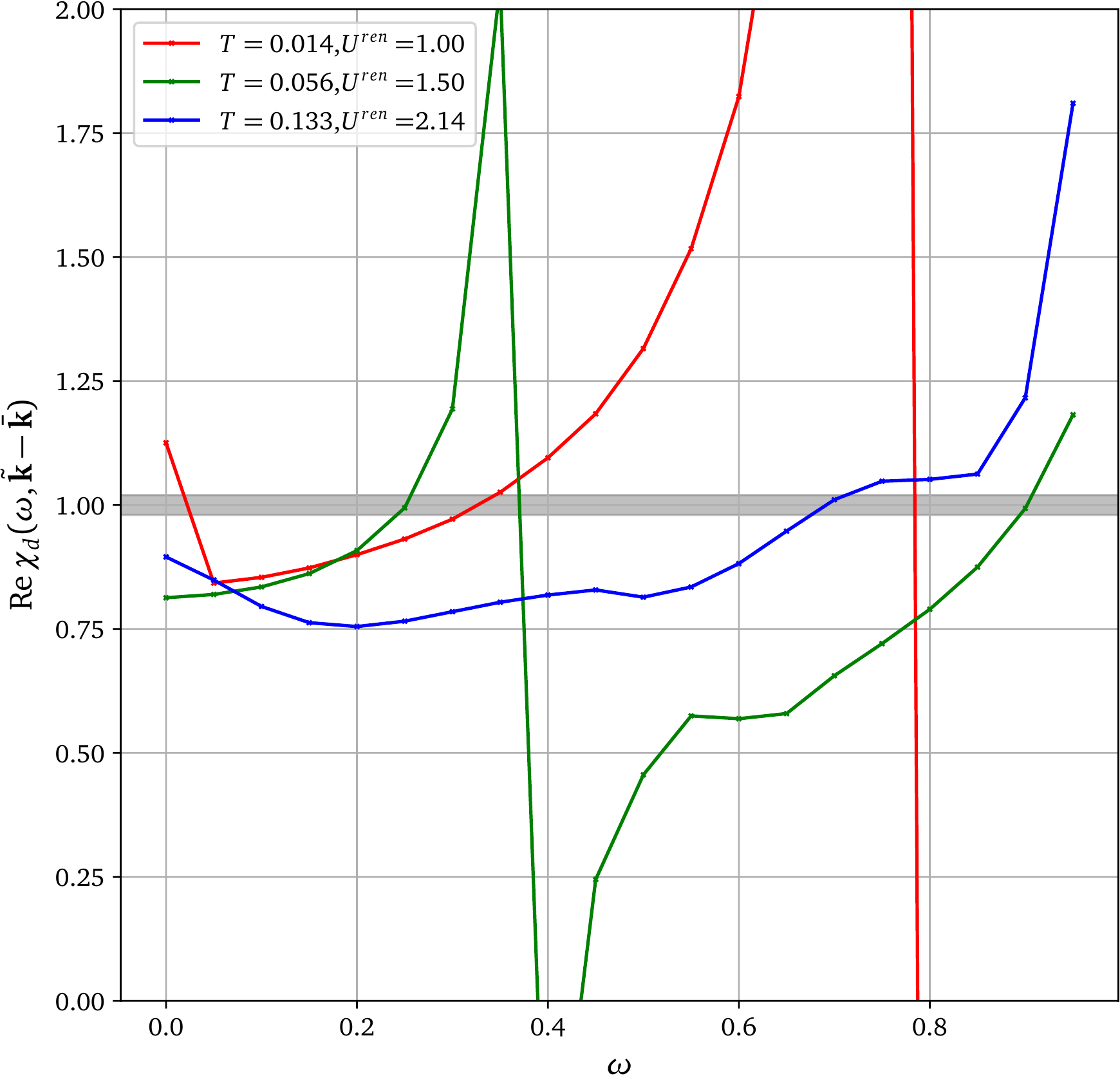}
    \caption{Real part of Eq.~\eqref{eq:discussion:analytical_continuation_denom} as a function of energy $\omega$. The energies at which the real part approaches $1$ correspond to 
    the peaks observed near $\omega\simeq 0$ and $\omega\in [0.5,0.8]$.
    }
    \label{fig:results:1D:poles_denom_single_ladder}
\end{figure}

\section{Discussion}
\label{sec:discussion}

To understand the origin of the $\pi$-ton peaks appearing in the energy range $\omega\in \left[0.5,0.8\right]$  (c.f.~Figs.~\ref{fig:results:1D:IPT:jj} and \ref{fig:results:1D:IPT:szsz}), we separately consider the numerator and denominator of Eq.~\eqref{eq:correlation_functions:single_ladder_k_space_representation_sus_corr}. We focus on Eq.~\eqref{eq:correlation_functions:single_ladder_k_space_representation_sus_corr} since for weak interactions, 
Eq.~\eqref{eq:correlation_functions:infinite_ladder_k_space_representation_sus_corr} yields only a small correction to the peak location.  The analytical continuation to the real-frequency axis of $\chi_d$ in the denominator of Eq.~\eqref{eq:correlation_functions:single_ladder_k_space_representation_sus_corr} reads
\begin{align}
\label{eq:discussion:analytical_continuation_denom}
&\chi^{\sigma,-\sigma}_d(\omega,\tilde{\mathbf{k}}-\bar{\mathbf{k}}) = U\int_{-\pi}^{\pi}\frac{\mathrm{d}^Dk}{(2\pi)^D}\iint_{-\infty}^{\infty}\mathrm{d}\omega^{\prime}\mathrm{d}\omega^{\prime\prime}\notag\\
&\hspace{12mm}\times\mathcal{A}_{\mathbf{k}+\tilde{\mathbf{k}}-\bar{\mathbf{k}}}(\omega^{\prime})\mathcal{A}_{\mathbf{k}}(\omega^{\prime\prime})\frac{n_F(\omega^{\prime})-n_F(\omega^{\prime\prime})}{\omega+i\eta-\left(\omega^{\prime}-\omega^{\prime\prime}\right)},
\end{align} 
where $\mathcal{A}_{\mathbf{k}}(\omega)=-\frac{1}{\pi}\operatorname{Im}\mathcal{G}(\mathbf{k},\omega)$ is the spectral function, $n_F$ is the Fermi-Dirac distribution, and $\eta\to 0^+$.

As mentioned before, the $\pi$-ton type vertex correction originates mainly from $\tilde{\mathbf{k}}-\bar{\mathbf{k}}=\pi$.
However, in Eq.~\eqref{eq:correlation_functions:single_ladder_k_space_representation_sus_corr} not all such $(\tilde{\mathbf{k}},\bar{\mathbf{k}})$-tuples give comparable contributions when one considers the numerator together with the denominator. The reason is that the $\mathbf{k}$-dependent spectral weight of the interacting Green's function varies with momentum: the spectral functions for momentum values around $\mathbf{k}=\pm\frac{\pi}{2}$ exhibit a sharp peak near $\omega=0$, 
while the spectra broaden as $\mathbf{k}$ approaches $0$ or $\pi$ (corresponding to peak positions near $\omega=\pm 2$). This means that the numerator of Eqs.~\eqref{eq:correlation_functions:single_ladder_k_space_representation_sus_corr} and \eqref{eq:correlation_functions:infinite_ladder_k_space_representation_sus_corr} yields the largest values for tuples $(\tilde{\mathbf{k}}\simeq\pm\frac{\pi}{2},\bar{\mathbf{k}}\simeq\mp\frac{\pi}{2})$. On the other hand, the real part of the denominator of Eq.~\eqref{eq:discussion:analytical_continuation_denom} approaches $1$ in the energy range $\omega\in \left[0.4,0.9\right]$ for all considered interaction values, as illustrated in Fig.~\ref{fig:results:1D:poles_denom_single_ladder}. This energy range corresponds to the $\mathcal{A}_{\mathbf{k}}$ peak position near $\mathbf{k}=\pm\frac{\pi}{2}$, so that the corresponding poles get amplified by the numerator and show up as peaks in the responses. Note that, as shown in Fig.~\ref{fig:results:1D:poles_denom_single_ladder}, for all values of the interaction except $U=3$ ($U^\text{ren}=2.14$), a second pole appears even closer to $\omega=0$ implying a rise in the responses close to $\omega=0$ (c.f.~Figs.~\ref{fig:results:1D:IPT:jj} and \ref{fig:results:1D:IPT:szsz}). This is the origin of the observed broadening of the Drude peak.


\section{Conclusions}
\label{sec:conclusion}

We have studied the effect of $\pi$-ton-ladder-type vertex corrections on the $q=0$ optical and magnetic response of the half-filled Hubbard model. This type of vertex correction has been identified in Ref.~\onlinecite{kauch_pitons_2019} as the most relevant one in the vicinity of an ordered phase with ordering wave vector ${\bf k}_\pi$, as in the present system near the AFM phase boundary. We have considered RPA-$\pi$-tons, where the vertical ladder is constructed using a (properly renormalized) instantaneous Hubbard interaction, instead of a vertex, and with interacting Green's functions obtained from a DMFT simulation. According to the results of Kauch {\it et al.}\cite{kauch_pitons_2019} this should be a meaningful and efficient post-processing procedure which allows to incorporate relevant fluctuations into the $q=0$ responses measured with DMFT. 

This weak-coupling diagrammatic approach yields stable and physically plausible results for weak interactions, while the calculation of RPA-$\pi$-tons in the intermediate coupling and Mott regimes suffers from inconsistencies. In particular, in the Mott regime, the corresponding vertex correction is larger in the high temperature region than close to the AFM phase boundary (if the diagrams are evaluated with the more reliable NCA Green's functions). As a side remark, we note that also diagrammatic extensions of DMFT,\cite{RevModPhys.90.025003} even though more accurate, suffer convergence problems at intermediate to strong coupling, related to the multivaluedness of the Luttinger-Ward functional, or the occurrence of divergences in the two-particle-irreducible vertex functions.\cite{PhysRevLett.119.056402,PhysRevB.97.245136}

Even at weak interactions, the $U$ used in the evaluation of the $\pi$-ton diagram needs to be renormalized in order to shift the pole to the actual phase boundary. If this is done, the $\pi$-ton results at weak $U$ in a broadening of the Drude peak and in a characteristic in-gap feature of the optical and spin response. With increasing $U$, the Drude peak is suppressed, while the in-gap feature grows and shifts up in energy, eventually merging with the higher energy spectral weight associated with the bubble contribution. 

We have also considered a double-ladder extension of the RPA-$\pi$-ton to estimate the dominance of the single-ladder contribution. This double-ladder has little qualitative effect for $U\le 2$, and results mainly in a broadening and small red-shift of the $\pi$-ton peak and a suppression of the high-energy spectral weight in the optical conductivity. This indicates that the prominent features in the responses stem mainly from the single-ladder contributions to the vertex corrections. 

With increasing $U$, the poles of the ladder-type vertex corrections stray away from the DMFT phase boundary. This indicates that the contribution from other diagram topologies becomes significant and resorting solely to ladder-type vertex corrections (Eqs.~\eqref{eq:correlation_functions:single_ladder_k_space_representation_sus_corr} and \eqref{eq:correlation_functions:infinite_ladder_k_space_representation_sus_corr}) becomes insufficient to account for the relevant physics.

In the context of superconductivity, the vertex correction analogous to the single-ladder $\pi$-ton is known as the Maki-Thompson (MT) diagram.~\cite{10.1143/PTP.40.193,PhysRevB.1.327} In this community, the significance of the MT and other diagrams has been extensively discussed, and it is known that the Aslamazov-Larkin (AL) diagram,~\cite{Aslamazov1968EffectOF} with a topology different from both our single ladder and the double-ladder, plays an important role. Hence, also in the present context of optical and spin responses near an AFM phase, it may be important to also consider non-$\pi$-ton diagram topologies, including AL-type vertex corrections. Diagrammatic Monte Carlo \cite{Prokofev_2004,Vanhoucke_2010} would be an unbiased numerical technique to check the relevance of different classes of diagrams.      

Benchmarking our DMFT and RPA based post-processing scheme is difficult, because of a lack of numerically exact results for the optical conductivity and spin response. Our results for small $U$ are qualitatively consistent with the findings reported in Ref.~\onlinecite{kauch_pitons_2019} which were also obtained with approximate (but more advanced) formalisms. In particular, Kauch {\it et al.}, using different models, found similar broadenings of the Drude peak, in-gap features, and shifts in the edge of the high-energy spectral weight, which were traced back to the $\pi$-ton contribution. In order to benchmark against numerically exact data, we considered the 1D Hubbard model at $T=0$, which can be solved with DMRG. The physics of this system is however qualitatively different from that captured by the DMFT treatment, which is representative of high-dimensional finite-temperature systems, even if the 1D density of states is used in the self-consistency loop. One reason is the presence of a Mott gap at $T=0$ in the half-filled 1D Hubbard model for any $U>0$, and the very asymmetric spin and charge response. While a $\pi$-ton contribution to the optical conductivity of the 1D system cannot be excluded, it is difficult to disentangle it from the dominant charge excitation peak. Comparisons to finite-temperature DMRG data would be an interesting topic for future investigations.

\begin{acknowledgments}
We thank A.-M. S. Tremblay, N. Tsuji, M. Eckstein, R. Nourafkan and M. Charlebois for helpful discussions. This work has been supported by SNSF Grant No. 200021-165539 and by JST CREST Grant No. JPMJCR19T3. The calculations have been performed on the Beo05 cluster at the University of Fribourg.    
\end{acknowledgments}


\appendix


\section{Derivation of the vertex corrections}
\label{sec:appendice:Derivation_vertex_corrections}

\begin{figure}[b]
  \centering
    \includegraphics[width=\linewidth]{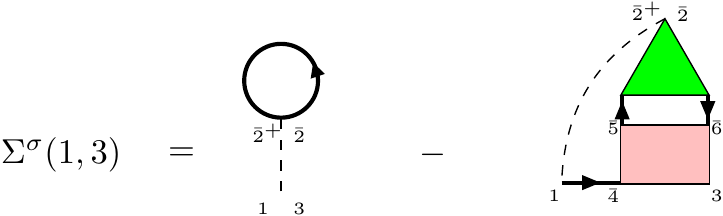}
      \caption{Diagrammatic representation of $\Sigma$ (Eq.~\eqref{eq:correlation_functions:self_energy}). 
      The first diagram is the Hartree diagram (tadpole) while the second term 
      contains an infinite number of diagrams that can be generated self-consistently. In the last term, both the green and pink shapes together define the vertex 
      function.~\cite{stefanucci_van_leeuwen_2013}
      }
  \label{fig:self_energy_diagrams}
\end{figure}

The equations of motion in $D$ dimension(s) read
\begin{align}
\label{eq:appendice:Derivation_vertex_corrections:differentiation_greens_function_for_equation_motion}
-&\partial_{\tau_1}\langle\mathcal{T}_{\tau}\hat{c}_{\sigma}(1)\hat{c}_{\sigma}^{\dagger}(\bar{2})\rangle_{\phi}=\notag\\ 
&-\delta^D(x_1-x_2)\delta(\tau_1-\tau_2) - \left<\mathcal{T}_{\tau}\partial_{\tau_1}\hat{c}_{\sigma}(1)\hat{c}_{\sigma}^{\dagger}(2)\right>_{\phi} -\notag\\
&\left<\mathcal{T}_{\tau}\left[\hat{K},\hat{c}_{\sigma}(1)\right]\hat{c}_{\sigma}^{\dagger}(2)\right>_{\phi},
\end{align} 
which, using Dyson's equation, yields
\begin{align}
\label{eq:appendice:Derivation_vertex_corrections:eq_motion_Hubbard_model}
&\Sigma_{\sigma}^{\phi}(1,\bar{2})\mathcal{G}_{\sigma}^{\phi}(\bar{2},2)=\notag\\
&\hspace{5mm}-\sum_{\sigma_2}
U\delta_{\sigma_2,-\sigma}\delta(1-\bar{2})
\left<\mathcal{T}_{\tau}\hat{c}_{\sigma}(1)\hat{c}^{\dagger}_{\sigma}(2)\hat{c}^{\dagger}_{\sigma_2}(\bar{2}^+)\hat{c}_{\sigma_2}(\bar{2})\right>_{\phi}.
\end{align} 
Making use of Eqs.~\eqref{eq:correlation_functions:four_point_correlation_function} and \eqref{eq:correlation_functions:identity_relation}, we may re-express the right-hand side as
\begin{align}
\label{eq:appendice:Derivation_vertex_corrections:self_energy_Hubbard_model}
&\Sigma_{\sigma}^{\phi}(1,\bar{2})\mathcal{G}_{\sigma}^{\phi}(\bar{2},2)=\notag\\
&-\sum_{\sigma_2}U\delta_{\sigma_2,-\sigma}\delta(1-\bar{2})
\biggl[\mathcal{G}^{\phi}_{\sigma}(1,\bar{2}^+)\mathcal{G}^{\phi}_{\sigma}(\bar{2},2)\delta_{\sigma,\sigma_2}-\notag\\
&\mathcal{G}^{\phi}_{\sigma_2}(\bar{2},\bar{2}^+)\mathcal{G}^{\phi}_{\sigma}(1,2) + \sum_{\sigma^{\prime}}\mathcal{G}^{\phi}_{\sigma}(1,\bar{4})\frac{\delta\Sigma^{\phi}_{\sigma}(\bar{4},\bar{3})}{\delta\mathcal{G}^{\phi}_{\sigma^{\prime}}(\bar{5},\bar{6})}\times\notag\\
&\frac{\delta\mathcal{G}^{\phi}_{\sigma^{\prime}}(\bar{5},\bar{6})}{\delta\phi_{\sigma_2}(\bar{2}^+,\bar{2})}\mathcal{G}^{\phi}_{\sigma}(\bar{3},2)\biggr].
\end{align} After some manipulations, one obtains the expression \eqref{eq:correlation_functions:self_energy} for the self-energy. The general diagrammatic form of this self-energy is shown in Fig.~\ref{fig:self_energy_diagrams}.

\begin{figure}[t]
  \centering
    \includegraphics[width=\linewidth]{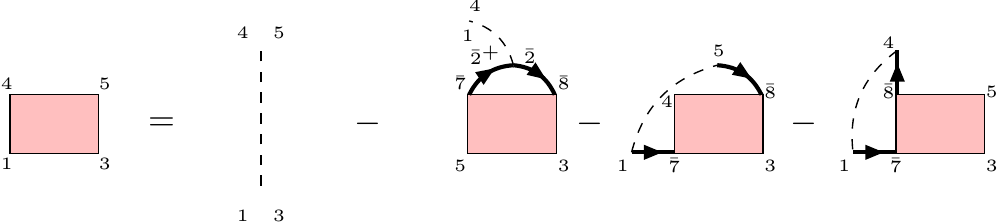}
      \caption{Diagrammatic representation of $\frac{\delta\Sigma}{\delta\mathcal{G}}$ when only keeping the first term of $\frac{\delta\mathcal{G}}{\delta\phi}$. The first and last terms are 
      those that will be retained in our approximation for $\frac{\delta\Sigma}{\delta\mathcal{G}}$.}
  \label{fig:delta_sigma_delta_G_Hubbard}
\end{figure}

To compute the susceptibilities introduced in Sec.~\ref{sec:correlation_functions}, one must work out expressions for $\frac{\delta\Sigma}{\delta\mathcal{G}}$ from Eq.~\eqref{eq:correlation_functions:self_energy}. Carrying out the functional derivative with respect to the interacting Green's function and keeping the first terms, one gets
\begin{align}
\label{eq:appendice:Derivation_vertex_corrections:delta_sigma_over_G_Hubbard}
&\frac{\delta\Sigma^{\phi}_{\sigma}(1,3)}{\delta\mathcal{G}^{\phi}_{\sigma^{\prime}}(4,5)}=U\delta_{\sigma^{\prime},-\sigma}\delta(4-5)\delta(1-3)\delta(1-4)  \notag\\
&\phantom{=}-U\delta(1-4)\delta_{\sigma,\sigma^{\prime}}\frac{\delta\Sigma^{\phi}_{\sigma}(5,3)}{\delta\mathcal{G}^{\phi}_{-\sigma}(\bar{7},\bar{8})}\mathcal{G}^{\phi}_{-\sigma}(\bar{7},1^+)\mathcal{G}^{\phi}_{-\sigma}(1,\bar{8})  \notag\\
&\phantom{=}-U\delta_{\sigma^{\prime},-\sigma}\delta(1-5)\mathcal{G}^{\phi}_{\sigma}(1,\bar{7})\frac{\delta\Sigma^{\phi}_{\sigma}(\bar{7},3)}{\delta\mathcal{G}^{\phi}_{\sigma^{\prime}}(4,\bar{8})}\mathcal{G}^{\phi}_{\sigma^{\prime}}(5,\bar{8})  \notag\\
&\phantom{=}-U\delta_{\sigma^{\prime},-\sigma}\delta(1-4)\mathcal{G}^{\phi}_{\sigma}(1,\bar{7})\frac{\delta\Sigma^{\phi}_{\sigma}(\bar{7},3)}{\delta\mathcal{G}^{\phi}_{\sigma^{\prime}}(\bar{8},5)}\mathcal{G}^{\phi}_{\sigma^{\prime}}(\bar{8},4^+) - \cdots.
\end{align} 
which is shown diagrammatically in Fig.~\ref{fig:delta_sigma_delta_G_Hubbard}.

The expression for a single vertical ladder is obtained when keeping the first and last terms in Eq.~\eqref{eq:appendice:Derivation_vertex_corrections:delta_sigma_over_G_Hubbard}. Isolating $\frac{\delta\Sigma}{\delta\mathcal{G}}$ from those two kept terms gives
\begin{align}
\label{eq:appendice:Derivation_vertex_corrections:functional_derivative_dself_energy_dG}
&\frac{\delta\Sigma_{\sigma}(1,3)}{\delta\mathcal{G}_{\sigma^{\prime}}(4,5)} =\notag\\
&\phantom{=}\frac{U\delta_{\sigma^{\prime},-\sigma}\delta(3-5)}{\delta(1-3)\delta(4-5)+U\delta(1-4)\mathcal{G}_{\sigma}(1,3)\mathcal{G}_{\sigma^{\prime}}(5,4^{+})}.
\end{align} This relation is the real-space equivalent of the $\square$ term Eq.~\eqref{eq:correlation_functions:single_ladder_k_space_square_expr}. The source field is set to 0 ($\phi\to 0$) in Eq.~\eqref{eq:appendice:Derivation_vertex_corrections:functional_derivative_dself_energy_dG} since this is the final form sought.

We next derive the expression for $\blacktriangleright^{(\text{even})}$, which sums up all even-ladder corrections to the triangle vertex:
\begin{widetext}
\begin{align}
\label{eq:appendice:Derivation_vertex_corrections:infinite_ladder_full_dG_dphi_even}
&\frac{\delta\mathcal{G}^{\phi,(\text{even})}_{\sigma^{\prime\prime}}(5,6)}{\delta\phi_{\sigma^{\prime}}(2^+,2)} =\notag\\
& \mathcal{G}^{\phi}_{\sigma^{\prime\prime}}(5,2^+)\mathcal{G}^{\phi}_{\sigma^{\prime\prime}}(2,6)\delta_{\sigma^{\prime\prime},\sigma^{\prime}}+\sum_{\tilde{\sigma}^{\prime},\tilde{\sigma}^{\prime\prime}}\mathcal{G}^{\phi}_{\sigma^{\prime\prime}}(5,\bar{1})\mathcal{G}^{\phi}_{\sigma^{\prime\prime}}(\bar{3},6)\frac{\delta\Sigma^{\phi}_{\sigma^{\prime\prime}}(\bar{1},\bar{3})}{\delta\mathcal{G}^{\phi}_{\tilde{\sigma}^{\prime}}(\bar{7},\bar{8})}\mathcal{G}^{\phi}_{\tilde{\sigma}^{\prime}}(\bar{7},\bar{10})\mathcal{G}^{\phi}_{\tilde{\sigma}^{\prime}}(\bar{11},\bar{8})\frac{\delta\Sigma^{\phi}_{\tilde{\sigma}^{\prime}}(\bar{10},\bar{11})}{\delta\mathcal{G}^{\phi}_{\tilde{\sigma}^{\prime\prime}}(\bar{12},\bar{13})}\frac{\delta\mathcal{G}^{\phi,(\text{even})}_{\tilde{\sigma}^{\prime\prime}}(\bar{12},\bar{13})}{\delta\phi_{\sigma^{\prime}}(2^+,2)}\notag\\
&\Leftrightarrow\notag\\
&\blacktriangleright_{\sigma^{\prime\prime}\sigma^{\prime}}^{\phi,(\text{even})}(5,6,2) =\notag\\
&\mathcal{G}^{\phi}_{\sigma^{\prime\prime}}(5,2^+)\mathcal{G}^{\phi}_{\sigma^{\prime\prime}}(2,6)\delta_{\sigma^{\prime\prime},\sigma^{\prime}}+\underbrace{\sum_{\tilde{\sigma}^{\prime},\tilde{\sigma}^{\prime\prime}}\mathcal{G}^{\phi}_{\sigma^{\prime\prime}}(5,\bar{7})\mathcal{G}^{\phi}_{\sigma^{\prime\prime}}(\bar{8},6)\square^{\phi}_{\sigma^{\prime\prime}\tilde{\sigma}^{\prime}}(\bar{7}-\bar{8})\mathcal{G}^{\phi}_{\tilde{\sigma}^{\prime}}(\bar{7},\bar{10})\mathcal{G}^{\phi}_{\tilde{\sigma}^{\prime}}(\bar{11},\bar{8})\square^{\phi}_{\tilde{\sigma}^{\prime}\tilde{\sigma}^{\prime\prime}}(\bar{10}-\bar{11})\blacktriangleright_{\tilde{\sigma}^{\prime\prime}\sigma^{\prime}}^{\phi,(\text{even})}(\bar{10},\bar{11},2)}_{\equiv \ \blacktriangleright^{\phi,(\text{even}),\text{corr}}_{\sigma^{\prime\prime}\sigma^{\prime}}}.
\end{align}
\end{widetext} 

\begin{figure}[ht]
  \centering
    \includegraphics[width=\linewidth]{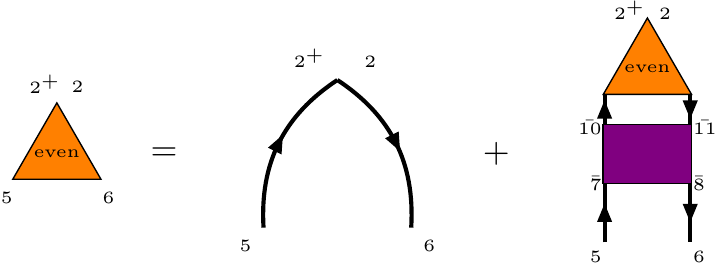}
      \caption{Diagrammatic representation of the terms of $\blacktriangleright^{(\text{even})}$ containing solely an even number of vertical ladders (pink boxes); the violet box is 
      comprised of two pink boxes connected together by two Green's functions. The first term is the one that will be retained in our approximation.}
  \label{fig:delta_G_delta_phi_diagrams_even}
\end{figure}

Eq.~\eqref{eq:appendice:Derivation_vertex_corrections:infinite_ladder_full_dG_dphi_even} is illustrated diagrammatically in Fig.~\ref{fig:delta_G_delta_phi_diagrams_even}. Note that self-consistently substituting $\blacktriangleright^{(\text{even})}$ into the right hand side will generate a ladder containing an even number of vertical ladders ($\square$). $\blacktriangleright^{(\text{even}),\text{corr}}$
corresponds to the last term in Fig.~\ref{fig:delta_G_delta_phi_diagrams_even}, where the two ladders are represented by a violet box. Retaining only the first term on the right hand side of \eqref{eq:appendice:Derivation_vertex_corrections:infinite_ladder_full_dG_dphi_even}, one obtains the real-space expression for the single-ladder vertex corrections, corresponding to the lowest-order vertex correction consisting of an odd number of vertical ladders:

\begin{align}
\label{eq:appendice:Derivation_vertex_corrections:truncated_delta_G_phi_real_space_representation_susceptibility}
&\chi_{\text{sl}}^{\sigma\sigma^{\prime}}(1,2) = \notag\\
&-\frac{U\delta_{\sigma^{\prime},-\sigma}\mathcal{G}_{\sigma}(1,\bar{5})\mathcal{G}_{\sigma^{\prime}}(\bar{6},2^{+})\mathcal{G}_{\sigma^{\prime}}(2,\bar{3})\mathcal{G}_{\sigma}(\bar{3},1^{+})}{\delta(\bar{5}-\bar{3})\delta(\bar{6}-\bar{3}) + U\delta(\bar{5}-\bar{6})\mathcal{G}_{\sigma}(\bar{5},\bar{3})\mathcal{G}_{\sigma^{\prime}}(\bar{3},\bar{6})},
\end{align} with $\phi\to 0$.
Fourier transformed to ($\mathbf{k},\omega_n$) space, Eq.~\eqref{eq:appendice:Derivation_vertex_corrections:truncated_delta_G_phi_real_space_representation_susceptibility} gives Eq.~\eqref{eq:correlation_functions:single_ladder_k_space_representation_sus_corr}.

Similarly to $\blacktriangleright^{(\text{even})}$, one can come up with an expression for $\blacktriangleright^{(\text{odd})}$ to calculate the double-ladder ($\chi_{\text{dl}}$) and higher-order even-ladder corrections. The only difference between $\blacktriangleright^{(\text{odd})}$ and $\blacktriangleright^{(\text{even})}$ is the first term with which the diagrams are generated self-consistently. Therefore, considering the notation introduced and the symmetries inherited from the Hubbard model, one gets

\begin{figure}[ht]
  \centering
    \includegraphics[width=\linewidth]{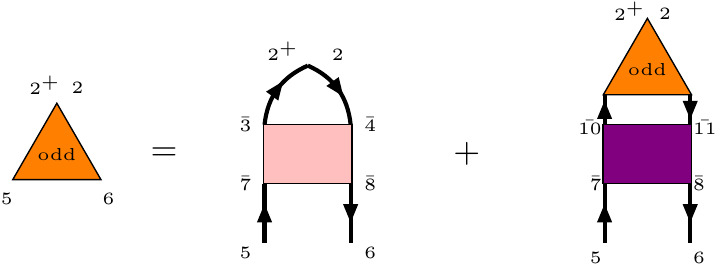}
      \caption{Diagrammatic representation of the terms of $\blacktriangleright^{(\text{odd})}$ containing solely an odd number of vertical ladders (pink box); again the violet box is comprised of two pink boxes connected together by two 
      Green's functions. The first term is the one that will be retained in our approximation. Note that in the case of the Hubbard model, for instantaneous interaction, $\bar{7}=\bar{3}$ and $\bar{8}=\bar{4}$ for the pink box.}
  \label{fig:delta_G_delta_phi_diagrams_odd}
\end{figure}

\begin{widetext}
\begin{align}
\label{eq:appendice:Derivation_vertex_corrections:infinite_ladder_full_dG_dphi_odd}
\blacktriangleright_{\sigma^{\prime\prime}\sigma^{\prime}}^{\phi,(\text{odd})}(5,6,2) = \mathcal{G}^{\phi}_{\sigma^{\prime\prime}}(5,\bar{7})\mathcal{G}^{\phi}_{\sigma^{\prime\prime}}(\bar{8},6)\square^{\phi}_{\sigma^{\prime\prime}\sigma^{\prime}}(\bar{7}-\bar{8})\mathcal{G}^{\phi}_{\sigma^{\prime}}(\bar{7},2^+)\mathcal{G}^{\phi}_{\sigma^{\prime}}(2,\bar{8})+\blacktriangleright^{\phi,(\text{odd}),\text{corr}}_{\sigma^{\prime\prime}\sigma^{\prime}}(5,6,2).
\end{align}
Eq.~\eqref{eq:appendice:Derivation_vertex_corrections:infinite_ladder_full_dG_dphi_odd} has the diagrammatic representation shown in Fig.~\ref{fig:delta_G_delta_phi_diagrams_odd}. This time, Eq.~\eqref{eq:appendice:Derivation_vertex_corrections:infinite_ladder_full_dG_dphi_odd} generates iteratively terms consisting of a ladder containing an even number of vertical ladders. Keeping only the first-order term of Eq.~\eqref{eq:appendice:Derivation_vertex_corrections:infinite_ladder_full_dG_dphi_odd} and inserting it into Eq.~\eqref{eq:correlation_functions:general_charge_susceptibility_notation_corr} yields the expression for the double-ladder vertex corrections ($\phi\to 0$):

\begin{align}
\label{eq:appendice:ladder_of_ladders_real_space}
\chi_{\text{dl}}^{\sigma\sigma^{\prime}}(1,2) = -\sum_{\sigma^{\prime\prime}}\mathcal{G}_{\sigma}(1,\bar{4})\mathcal{G}_{\sigma}(\bar{3},1^{+})\square_{\sigma\sigma^{\prime\prime}}(\bar{4}-\bar{3})\mathcal{G}_{\sigma^{\prime\prime}}(\bar{4},\bar{7})\mathcal{G}_{\sigma^{\prime\prime}}(\bar{8},\bar{3})\square_{\sigma^{\prime\prime}\sigma^{\prime}}(\bar{7}-\bar{8})\mathcal{G}_{\sigma^{\prime}}(\bar{7},2^+)\mathcal{G}_{\sigma^{\prime}}(2,\bar{8}).
\end{align}
\end{widetext} 
Carrying out the Fourier transformation of Eq.~\eqref{eq:appendice:ladder_of_ladders_real_space} into ($\mathbf{k},\omega_n$) space yields Eq.~\eqref{eq:correlation_functions:infinite_ladder_k_space_representation_sus_corr}.

\section{Longitudinal conductivity}
\label{appendix:electronic_conductivity}

The continuity equation associated with the electric charge conservation reads

\begin{align}
\label{eq:appendix:electronic_conductivity:continuity_relation}
\frac{\partial\rho(\mathbf{r},t)}{\partial t} + \mathbf{\nabla}\cdot\mathbf{j}(\mathbf{r},t) = 0,
\end{align} where $\mathbf{j}$ is the current density, and $\rho$ is the charge density. When transforming Eq.~\eqref{eq:appendix:electronic_conductivity:continuity_relation} to Fourier space, one gets

\begin{align}
\label{eq:appendix:electronic_conductivity:continuity_relation_Fourier_space}
-\omega\rho(\mathbf{q},\omega) + \mathbf{q}\cdot\mathbf{j}(\mathbf{q},\omega) = 0.
\end{align} The two-particle spectral function $\chi^{\prime\prime}_{\rho\rho}$ corresponding to the charge density observable is, within the linear response theory framework,

\begin{align}
\label{eq:appendix:electronic_conductivity:linear_response_charge_density}
\chi^{\prime\prime}_{\rho\rho}(\mathbf{q},\omega) = \frac{1}{N}\langle\left[\hat{\rho}(\mathbf{q},\omega),\hat{\rho}(-\mathbf{q},-\omega)\right]\rangle_{\hat{\mathcal{H}}_0},
\end{align} where $\hat{\mathcal{H}}_0$ is the non-interacting Hamiltonian and $N$ is the $\mathbf{k}$-space grid, such that according to Eq.~\eqref{eq:appendix:electronic_conductivity:continuity_relation_Fourier_space}, the current-current two-body spectral function reads

\begin{align}
\label{eq:appendix:electronic_conductivity:linear_response_current_density}
\chi^{\prime\prime}_{j_ij_i}(q_i,\omega) = \frac{\omega^2}{q_i^2}\chi^{\prime\prime}_{\rho\rho}(q_i,\omega),
\end{align} where the index $i$ denotes the Cartesian axes. The spectral representation of the current-current correlation function $\chi_{j_ij_i}$ is

\begin{align}
\label{eq:appendix:electronic_conductivity:linear_response_current_density_corr_function}
\chi_{j_ij_i}(q_i,\omega) = \int\frac{\mathrm{d}\omega^{\prime}}{\pi}\frac{\chi_{j_ij_i}^{\prime\prime}(q_i,\omega^{\prime})}{\omega^{\prime}-\omega-i\eta}.
\end{align} 
With these ingredients we can derive an expression for the real part of the longitudinal electric conductivity, denoted $\operatorname{Re}\sigma_{ii}(q_i,\omega)$. According to  Eq.~\eqref{eq:appendix:electronic_conductivity:linear_response_current_density},

\begin{align}
\label{eq:appendix:electronic_conductivity:linear_response_current_density_key_rel}
&\int\frac{\mathrm{d}\omega}{\pi}\frac{\chi^{\prime\prime}_{j_ij_i}(q_i,\omega)}{\omega} = \frac{1}{q_i^2}\int\frac{\mathrm{d}\omega}{\pi}\omega\chi^{\prime\prime}_{\rho\rho}(q_i,\omega)\notag\\
&=\left[\frac{2i}{q_i^2}\frac{\partial}{\partial t}\int\frac{\mathrm{d}\omega}{2\pi}e^{-i\omega t}\chi_{\rho\rho}^{\prime\prime}(q_i,\omega)\right]\bigg\rvert_{t=0}\notag\\
&=\frac{1}{N q_i^2}\biggl<\left[i\frac{\partial}{\partial t}\hat{\rho}(q_i,t),\hat{\rho}(-q_i,0)\right]\bigg\rvert_{t=0}\biggr>_{\hat{\mathcal{H}}_0}\notag\\
&=\frac{1}{N q_i^2}\biggl<\left[\left[\hat{\rho}(q_i),\hat{\mathcal{H}}\right](t),\hat{\rho}(-q_i,0)\right]\bigg\rvert_{t=0}\biggr>_{\hat{\mathcal{H}}_0},
\end{align} where $\hat{\mathcal{H}}$ is the Hubbard Hamiltonian Eq.~\eqref{eq:Hubbard_model_intro}. The last expression of Eq.~\eqref{eq:appendix:electronic_conductivity:linear_response_current_density_key_rel} relates directly to the first moment of the density correlation function. Evaluating the commutators and taking the limit $q_i\to 0$, one gets

\begin{align}
\label{eq:appendix:electronic_conductivity:linear_response_current_density_key_rel_commutator_eval}
\int\frac{\mathrm{d}\omega}{\pi}\frac{\chi^{\prime\prime}_{j_ij_i}(q_i,\omega)}{\omega} = \frac{1}{q_i^2 N}\sum_{\mathbf{k},\sigma}\frac{\partial^2 \epsilon_{\mathbf{k},\sigma}}{\partial k_i^2}\langle \hat{n}_{\mathbf{k},\sigma}\rangle_{\mathcal{H}}\equiv -\langle\hat{j}_{i,d}\rangle,
\end{align} where $\epsilon_{\mathbf{k},\sigma}$ is the dispersion relation in $D$ dimensions and $\sigma$ is the spin. Eq.~\eqref{eq:appendix:electronic_conductivity:linear_response_current_density_key_rel_commutator_eval} will serve as a sum rule to verify if the longitudinal optical conductivity obeys conservation laws, namely $\operatorname{Re}\chi_{j_ij_i}(iq_n=0)=\sum_{\sigma}\idotsint_{-\pi}^{\pi}\mathrm{d}^{D}k\frac{\partial^2 \epsilon(\mathbf{k})}{\partial k_i}\langle n_{\mathbf{k},\sigma}\rangle$.  Equation~\eqref{eq:appendix:electronic_conductivity:linear_response_current_density_key_rel_commutator_eval} represents the diamagnetic contribution $\hat{j}_{i,d}$ to the current fluctuations $\delta\langle\hat{j}_i(\omega)\rangle$ (when multiplying the expression by $(-1)$), that is

\begin{align}
\label{eq:appendix:electronic_conductivity:linear_response_current_fluctuations}
\delta\langle\hat{j}_i(\omega)\rangle = \left[\langle\hat{j}_{i,d}\rangle+\chi_{j_ij_i}(\omega)\right]A_i(\omega).
\end{align} Hence, given that without scalar potential, the electric field obeys $E_i(t)=-\frac{\partial A_i(t)}{\partial t}$, the longitudinal conductivity reads

\begin{align}
\label{eq:appendix:electronic_conductivity:linear_response_longitudinal_conductivity}
\sigma_{ii}(q_i,\omega) = \frac{\langle\hat{j}_{i,d}\rangle+\chi_{j_ij_i}(\omega)}{i\left(\omega+i\eta\right)},
\end{align} owing to the relation linking the current fluctuations to the electric field in linear response theory: $\delta\langle j_i(\omega)\rangle = \sigma_{ii}(\omega)E_i(\omega)$. Now using Eqs.~\eqref{eq:appendix:electronic_conductivity:linear_response_current_density_corr_function} and \eqref{eq:appendix:electronic_conductivity:linear_response_current_density_key_rel_commutator_eval}, we aim at extracting the real part of the longitudinal conductivity:

\begin{align}
\label{eq:appendix:electronic_conductivity:linear_response_longitudinal_conductivity}
&\sigma_{ii}(q_i,\omega) =\notag\\
&\frac{1}{i\left(\omega + i\eta\right)}\left[\int\frac{\mathrm{d}\omega^{\prime}}{\pi}\frac{\chi_{j_ij_i}^{\prime\prime}(q_i,\omega^{\prime})}{\omega^{\prime}-\omega-i\eta}-\int\frac{\mathrm{d}\omega^{\prime}}{\pi}\frac{\chi^{\prime\prime}_{j_ij_i}(q_i,\omega^{\prime})}{\omega^{\prime}}\right]\notag\\
&= \frac{1}{i\left(\omega + i\eta\right)}\int\frac{\mathrm{d}\omega^{\prime}}{\pi}\frac{\left(\omega+i\eta\right)\chi_{j_ij_i}^{\prime\prime}(q_i,\omega^{\prime})}{\omega^{\prime}\left(\omega^{\prime}-\omega-i\eta\right)}\notag\\
&= \frac{1}{i}\int\frac{\mathrm{d}\omega^{\prime}}{\pi}\frac{\chi_{j_ij_i}^{\prime\prime}(q_i,\omega^{\prime})}{\omega^{\prime}\left(\omega^{\prime}-\omega-i\eta\right)}\notag\\
&\implies \operatorname{Re}\sigma_{ii}(q_i,\omega) = \frac{\chi_{j_ij_i}^{\prime\prime}(q_i,\omega)}{\omega}.
\end{align} 

\section{DMRG}
\label{appendix:DMRG}

Here, we briefly explain the basic principle of the  
DMRG method.~\cite{White1992PRL,Schollwock2011AnnPhys} In DMRG, quantum states are represented in the form of matrix product states (MPSs), 

\begin{align}
 \ket{\Psi}=&\sum_{\{\alpha_{i}\},\{s_{i}\}}
   M_{\alpha_{1}}[s_{1}]
   M_{\alpha_{1}\alpha_{2}}[s_{2}]
   M_{\alpha_{2}\alpha_{3}}[s_{3}]\cdots
   M_{\alpha_{N-1}}[s_{N}]\nonumber\\
   &\times
     \ket{s_{1},s_{2},\ldots,s_{N}},
\end{align}
where $N$ is the number of sites, 
$s_{i}$ represents the quantum state on site $i$, 
and in the present system $s_{i}=0,1,2,3$ correspond to 
$(n_{i\uparrow},n_{i\downarrow})=(0,0),(1,0),(0,1),(1,1)$, respectively 
($n_{i\uparrow}$ and $n_{i\downarrow}$ are the number of 
electrons with spin-up and spin-down). 
$\alpha_{i}$ ($i=1,\ldots,N-1$) 
is the suffix for the matrices. 
We also represent the Hamiltonian as 
a matrix product operator (MPO) 
\begin{align}
 \hat{\mathcal{H}}=&\sum_{\{\beta_{i}\},\{s_{i}\},\{s'_{i}\}}
   P_{\beta_{1}}[s_{1},s'_{1}]
   P_{\beta_{1}\beta_{2}}[s_{2},s'_{2}]\cdots
   P_{\beta_{N-1}}[s_{N},s'_{N}]\nonumber\\
   &\times
     \ket{s_{1},s_{2},\ldots,s_{N}}
     \bra{s'_{1},s'_{2},\ldots,s'_{N}}.
\end{align}
DMRG is a method to obtain the MPS of 
the ground state variationally using the MPO form 
of the Hamiltonian. 
The initial MPS is derived as follows. 
First, diagonalizing the two-site Hamiltonian 
\begin{align}
 \hat{\mathcal{H}}_{1}=&
   \sum_{\beta_{1},s_{1},s'_{1},s_{N},s'_{N}}
   P_{\beta_{1}}[s_{1},s'_{1}]
   P_{\beta_{1}}[s_{N},s'_{N}]
   \ket{s_{1},s_{N}}\bra{s'_{1},s'_{N}}
\end{align}
and performing the Schmidt decomposition 
to the lowest energy state, we obtain the matrices 
$M_{\alpha_{1}}[s_{1}]$ and $M_{\alpha_{N-1}}[s_{N}]$ 
(in the latter $\alpha_{1}$ is relabeled as $\alpha_{N-1}$). 
Next we can construct $M[s_{j+1}]$ and $M[s_{N-j}]$ 
from $M[s_{1}],\ldots,M[s_{j}]$ and 
$M[s_{N-j+1}],\ldots,M[s_{N}]$. 
We build the matrix 
\begin{widetext}
\begin{align}
 \hat{\mathcal{H}}_{j}=&
   \sum_{\{\beta_{i}\},\{s_{i}\},\{s'_{i}\}}
   \sum_{\substack{\alpha_{1},\ldots,\alpha_{j-1}\\
           \alpha_{N-j+1},\ldots,\alpha_{N}}}
   \sum_{\substack{\alpha'_{1},\ldots,\alpha'_{j-1}\\
           \alpha'_{N-j+1},\ldots,\alpha'_{N}}}
   M_{\alpha_{1}}^{*}[s_{1}]\cdots
     M_{\alpha_{j-1}\alpha_{j}}^{*}[s_{j}]
   M_{\alpha_{N-j}\alpha_{N-j+1}}^{*}[s_{N-j+1}]\cdots
     M_{\alpha_{N-1}}^{*}[s_{N}]\nonumber\\
   & \times M_{\alpha'_{1}}[s'_{1}]\cdots
     M_{\alpha'_{j-1}\alpha'_{j}}[s'_{j}]
   M_{\alpha'_{N-j}\alpha'_{N-j+1}}[s'_{N-j+1}]\cdots
     M_{\alpha'_{N-1}}[s'_{N}]
   P_{\beta_{1}}[s_{1},s'_{1}]\cdots
   P_{\beta_{j-1}\beta_{j}}[s_{j},s'_{j}]\nonumber\\
   & \times P_{\beta_{j}\beta_{j+1}}[s_{j+1},s'_{j+1}]
   P_{\beta_{j+1}\beta_{N-j}}[s_{N-j},s'_{N-j}]
   P_{\beta_{N-j}\beta_{N-j+1}}[s_{N-j+1},s'_{N-j+1}]\cdots
   P_{\beta_{N-1}}[s_{N},s'_{N}]
     \ket{s_{j+1},s_{N-j}}\bra{s'_{j+1},s'_{N-j}},
\nonumber
\end{align}
and calculate the lowest energy state by the Lanczos method. 
After the Schmidt decomposition, 
we only keep the bond indices corresponding to 
the $\chi$ largest singular values and truncate the others to obtain the matrices 
$M_{\alpha_{j}\alpha_{j+1}}[s_{j+1}]$ and 
$M_{\alpha_{N-j-1}\alpha_{N-j}}[s_{N-j}]$. 
In our study, we set $\chi=32$. 
By repeating this process $N/2-1$ times, 
the initial MPS $M[s_{1}],\ldots,M[s_{N}]$ is constructed. 

Then we optimize this MPS by the variational method. 
As for the two neighboring sites $j$ and $j+1$, 
we construct the matrix in a similar way as above 
\begin{align}
 &\hat{\mathcal{H}}_{\mathrm{var}}=
   \sum_{\{\beta_{i}\},\{s_{i}\},\{s'_{i}\}}
   \sum_{\substack{\alpha_{1},\ldots,\alpha_{j-2}\\
           \alpha_{j+2},\ldots,\alpha_{N}}}
   \sum_{\substack{\alpha'_{1},\ldots,\alpha'_{j-2}\\
           \alpha'_{j+2},\ldots,\alpha'_{N}}}
   M_{\alpha_{1}}^{*}[s_{1}]\cdots
   M_{\alpha_{j-2}\alpha_{j-1}}^{*}[s_{j-1}]
   M_{\alpha_{j+1}\alpha_{j+2}}^{*}[s_{j+2}]\cdots
   M_{\alpha_{N-1}}^{*}[s_{N}]\nonumber\\
   &\times M_{\alpha'_{1}}[s'_{1}]\cdots
   M_{\alpha'_{j-2}\alpha'_{j-1}}[s'_{j-1}]
   M_{\alpha'_{j+1}\alpha'_{j+2}}[s'_{j+2}]\cdots
   M_{\alpha'_{N-1}}[s'_{N}]
   P_{\beta_{1}}[s_{1},s'_{1}]
   P_{\beta_{1},\beta_{2}}[s_{2},s'_{2}]\cdots
   P_{\beta_{N-1}}[s_{N},s'_{N}]
   \ket{s_{j},s_{j+1}}\bra{s'_{j},s'_{j+1}},
\nonumber
\end{align}
\end{widetext}
and calculate the lowest energy state by the Lanczos method. 
Then the matrices on the sites $j$ and $j+1$, that is 
$M_{\alpha_{j-1}\alpha_{j}}[s_{j}]$ and 
$M_{\alpha_{j}\alpha_{j+1}}[s_{j+1}]$, 
can be updated by the Schmidt decomposition 
and the bond truncation. 
We perform this variational update process 
on pairs of neighboring sites, 
$(j,j+1),(j+1,j+2),(j+2,j+3),\ldots$ and
sweep over all sites iteratively 
until the calculated value of energy converges. 
Thus we can derive the MPS form of the ground state.

\bibliography{Bibliography}

\begin{thebibliography}{36}%
\makeatletter
\providecommand \@ifxundefined [1]{%
 \@ifx{#1\undefined}
}%
\providecommand \@ifnum [1]{%
 \ifnum #1\expandafter \@firstoftwo
 \else \expandafter \@secondoftwo
 \fi
}%
\providecommand \@ifx [1]{%
 \ifx #1\expandafter \@firstoftwo
 \else \expandafter \@secondoftwo
 \fi
}%
\providecommand \natexlab [1]{#1}%
\providecommand \enquote  [1]{``#1''}%
\providecommand \bibnamefont  [1]{#1}%
\providecommand \bibfnamefont [1]{#1}%
\providecommand \citenamefont [1]{#1}%
\providecommand \href@noop [0]{\@secondoftwo}%
\providecommand \href [0]{\begingroup \@sanitize@url \@href}%
\providecommand \@href[1]{\@@startlink{#1}\@@href}%
\providecommand \@@href[1]{\endgroup#1\@@endlink}%
\providecommand \@sanitize@url [0]{\catcode `\\12\catcode `\$12\catcode
  `\&12\catcode `\#12\catcode `\^12\catcode `\_12\catcode `\%12\relax}%
\providecommand \@@startlink[1]{}%
\providecommand \@@endlink[0]{}%
\providecommand \url  [0]{\begingroup\@sanitize@url \@url }%
\providecommand \@url [1]{\endgroup\@href {#1}{\urlprefix }}%
\providecommand \urlprefix  [0]{URL }%
\providecommand \Eprint [0]{\href }%
\providecommand \doibase [0]{http://dx.doi.org/}%
\providecommand \selectlanguage [0]{\@gobble}%
\providecommand \bibinfo  [0]{\@secondoftwo}%
\providecommand \bibfield  [0]{\@secondoftwo}%
\providecommand \translation [1]{[#1]}%
\providecommand \BibitemOpen [0]{}%
\providecommand \bibitemStop [0]{}%
\providecommand \bibitemNoStop [0]{.\EOS\space}%
\providecommand \EOS [0]{\spacefactor3000\relax}%
\providecommand \BibitemShut  [1]{\csname bibitem#1\endcsname}%
\let\auto@bib@innerbib\@empty
\bibitem [{\citenamefont {Uchida}\ \emph {et~al.}(1991)\citenamefont {Uchida},
  \citenamefont {Ido}, \citenamefont {Takagi}, \citenamefont {Arima},
  \citenamefont {Tokura},\ and\ \citenamefont {Tajima}}]{Uchida_1991}%
  \BibitemOpen
  \bibfield  {author} {\bibinfo {author} {\bibfnamefont {S.}~\bibnamefont
  {Uchida}}, \bibinfo {author} {\bibfnamefont {T.}~\bibnamefont {Ido}},
  \bibinfo {author} {\bibfnamefont {H.}~\bibnamefont {Takagi}}, \bibinfo
  {author} {\bibfnamefont {T.}~\bibnamefont {Arima}}, \bibinfo {author}
  {\bibfnamefont {Y.}~\bibnamefont {Tokura}}, \ and\ \bibinfo {author}
  {\bibfnamefont {S.}~\bibnamefont {Tajima}},\ }\href {\doibase
  10.1103/PhysRevB.43.7942} {\bibfield  {journal} {\bibinfo  {journal} {Phys.
  Rev. B}\ }\textbf {\bibinfo {volume} {43}},\ \bibinfo {pages} {7942}
  (\bibinfo {year} {1991})}\BibitemShut {NoStop}%
\bibitem [{\citenamefont {Basov}\ and\ \citenamefont
  {Timusk}(2005)}]{Basov_2005}%
  \BibitemOpen
  \bibfield  {author} {\bibinfo {author} {\bibfnamefont {D.~N.}\ \bibnamefont
  {Basov}}\ and\ \bibinfo {author} {\bibfnamefont {T.}~\bibnamefont {Timusk}},\
  }\href {\doibase 10.1103/RevModPhys.77.721} {\bibfield  {journal} {\bibinfo
  {journal} {Rev. Mod. Phys.}\ }\textbf {\bibinfo {volume} {77}},\ \bibinfo
  {pages} {721} (\bibinfo {year} {2005})}\BibitemShut {NoStop}%
\bibitem [{\citenamefont {Kauch}\ \emph {et~al.}(2020)\citenamefont {Kauch},
  \citenamefont {Pudleiner}, \citenamefont {Astleithner}, \citenamefont
  {Thunstr\"om}, \citenamefont {Ribic},\ and\ \citenamefont
  {Held}}]{kauch_pitons_2019}%
  \BibitemOpen
  \bibfield  {author} {\bibinfo {author} {\bibfnamefont {A.}~\bibnamefont
  {Kauch}}, \bibinfo {author} {\bibfnamefont {P.}~\bibnamefont {Pudleiner}},
  \bibinfo {author} {\bibfnamefont {K.}~\bibnamefont {Astleithner}}, \bibinfo
  {author} {\bibfnamefont {P.}~\bibnamefont {Thunstr\"om}}, \bibinfo {author}
  {\bibfnamefont {T.}~\bibnamefont {Ribic}}, \ and\ \bibinfo {author}
  {\bibfnamefont {K.}~\bibnamefont {Held}},\ }\href {\doibase
  10.1103/PhysRevLett.124.047401} {\bibfield  {journal} {\bibinfo  {journal}
  {Phys. Rev. Lett.}\ }\textbf {\bibinfo {volume} {124}},\ \bibinfo {pages}
  {047401} (\bibinfo {year} {2020})}\BibitemShut {NoStop}%
\bibitem [{\citenamefont {Georges}\ \emph {et~al.}(1996)\citenamefont
  {Georges}, \citenamefont {Kotliar}, \citenamefont {Krauth},\ and\
  \citenamefont {Rozenberg}}]{Georges_1996}%
  \BibitemOpen
  \bibfield  {author} {\bibinfo {author} {\bibfnamefont {A.}~\bibnamefont
  {Georges}}, \bibinfo {author} {\bibfnamefont {G.}~\bibnamefont {Kotliar}},
  \bibinfo {author} {\bibfnamefont {W.}~\bibnamefont {Krauth}}, \ and\ \bibinfo
  {author} {\bibfnamefont {M.~J.}\ \bibnamefont {Rozenberg}},\ }\href {\doibase
  10.1103/RevModPhys.68.13} {\bibfield  {journal} {\bibinfo  {journal} {Rev.
  Mod. Phys.}\ }\textbf {\bibinfo {volume} {68}},\ \bibinfo {pages} {13}
  (\bibinfo {year} {1996})}\BibitemShut {NoStop}%
\bibitem [{\citenamefont {White}(1992)}]{White1992PRL}%
  \BibitemOpen
  \bibfield  {author} {\bibinfo {author} {\bibfnamefont {S.~R.}\ \bibnamefont
  {White}},\ }\href {\doibase 10.1103/PhysRevLett.69.2863} {\bibfield
  {journal} {\bibinfo  {journal} {Phys. Rev. Lett.}\ }\textbf {\bibinfo
  {volume} {69}},\ \bibinfo {pages} {2863} (\bibinfo {year}
  {1992})}\BibitemShut {NoStop}%
\bibitem [{\citenamefont {Schollw\"ock}(2011)}]{Schollwock2011AnnPhys}%
  \BibitemOpen
  \bibfield  {author} {\bibinfo {author} {\bibfnamefont {U.}~\bibnamefont
  {Schollw\"ock}},\ }\href {\doibase https://doi.org/10.1016/j.aop.2010.09.012}
  {\bibfield  {journal} {\bibinfo  {journal} {Ann. Phys.}\ }\textbf {\bibinfo
  {volume} {326}},\ \bibinfo {pages} {96 } (\bibinfo {year}
  {2011})}\BibitemShut {NoStop}%
\bibitem [{\citenamefont {Arsenault}\ \emph {et~al.}(2012)\citenamefont
  {Arsenault}, \citenamefont {S\'emon},\ and\ \citenamefont
  {Tremblay}}]{arsenault_benchmark_2012}%
  \BibitemOpen
  \bibfield  {author} {\bibinfo {author} {\bibfnamefont {L.-F. m.~c.}\
  \bibnamefont {Arsenault}}, \bibinfo {author} {\bibfnamefont {P.}~\bibnamefont
  {S\'emon}}, \ and\ \bibinfo {author} {\bibfnamefont {A.-M.~S.}\ \bibnamefont
  {Tremblay}},\ }\href {\doibase 10.1103/PhysRevB.86.085133} {\bibfield
  {journal} {\bibinfo  {journal} {Phys. Rev. B}\ }\textbf {\bibinfo {volume}
  {86}},\ \bibinfo {pages} {085133} (\bibinfo {year} {2012})}\BibitemShut
  {NoStop}%
\bibitem [{\citenamefont {Kajueter}\ and\ \citenamefont
  {Kotliar}(1996)}]{kajueter_new_1996}%
  \BibitemOpen
  \bibfield  {author} {\bibinfo {author} {\bibfnamefont {H.}~\bibnamefont
  {Kajueter}}\ and\ \bibinfo {author} {\bibfnamefont {G.}~\bibnamefont
  {Kotliar}},\ }\href {\doibase 10.1103/PhysRevLett.77.131} {\bibfield
  {journal} {\bibinfo  {journal} {Phys. Rev. Lett.}\ }\textbf {\bibinfo
  {volume} {77}},\ \bibinfo {pages} {131} (\bibinfo {year} {1996})}\BibitemShut
  {NoStop}%
\bibitem [{\citenamefont {Grewe}\ and\ \citenamefont
  {Keiter}(1981)}]{PhysRevB.24.4420}%
  \BibitemOpen
  \bibfield  {author} {\bibinfo {author} {\bibfnamefont {N.}~\bibnamefont
  {Grewe}}\ and\ \bibinfo {author} {\bibfnamefont {H.}~\bibnamefont {Keiter}},\
  }\href {\doibase 10.1103/PhysRevB.24.4420} {\bibfield  {journal} {\bibinfo
  {journal} {Phys. Rev. B}\ }\textbf {\bibinfo {volume} {24}},\ \bibinfo
  {pages} {4420} (\bibinfo {year} {1981})}\BibitemShut {NoStop}%
\bibitem [{\citenamefont {Bickers}(1987)}]{RevModPhys.59.845}%
  \BibitemOpen
  \bibfield  {author} {\bibinfo {author} {\bibfnamefont {N.~E.}\ \bibnamefont
  {Bickers}},\ }\href {\doibase 10.1103/RevModPhys.59.845} {\bibfield
  {journal} {\bibinfo  {journal} {Rev. Mod. Phys.}\ }\textbf {\bibinfo {volume}
  {59}},\ \bibinfo {pages} {845} (\bibinfo {year} {1987})}\BibitemShut
  {NoStop}%
\bibitem [{\citenamefont {Bryan}(1990)}]{bryan_maximum_1990}%
  \BibitemOpen
  \bibfield  {author} {\bibinfo {author} {\bibfnamefont {R.~K.}\ \bibnamefont
  {Bryan}},\ }\href {\doibase 10.1007/BF02427376} {\bibfield  {journal}
  {\bibinfo  {journal} {Euro. Biophys. J.}\ }\textbf {\bibinfo {volume} {18}},\
  \bibinfo {pages} {165} (\bibinfo {year} {1990})}\BibitemShut {NoStop}%
\bibitem [{\citenamefont {Tsuji}\ and\ \citenamefont
  {Werner}(2013)}]{tsuji_nonequilibrium_2013}%
  \BibitemOpen
  \bibfield  {author} {\bibinfo {author} {\bibfnamefont {N.}~\bibnamefont
  {Tsuji}}\ and\ \bibinfo {author} {\bibfnamefont {P.}~\bibnamefont {Werner}},\
  }\href {\doibase 10.1103/PhysRevB.88.165115} {\bibfield  {journal} {\bibinfo
  {journal} {Phys. Rev. B}\ }\textbf {\bibinfo {volume} {88}},\ \bibinfo
  {pages} {165115} (\bibinfo {year} {2013})}\BibitemShut {NoStop}%
\bibitem [{Note1()}]{Note1}%
  \BibitemOpen
  \bibinfo {note} {See Appendix B of Ref.~\protect \rev@citealpnum
  {arsenault_benchmark_2012} and references therein for more details about the
  implementation.}\BibitemShut {Stop}%
\bibitem [{\citenamefont {Tremblay}(2017)}]{Tremblay:notes}%
  \BibitemOpen
  \bibfield  {author} {\bibinfo {author} {\bibfnamefont {A.-M.~S.}\
  \bibnamefont {Tremblay}},\ }\href
  {https://www.physique.usherbrooke.ca/tremblay/cours/phy-892/N-corps-2017.pdf}
  {\emph {\bibinfo {title} {{P}roblème à {N}-corps}}}\ (\bibinfo  {publisher}
  {Lecture notes},\ \bibinfo {address} {Sherbrooke, Qc, Canada},\ \bibinfo
  {year} {2017})\ \bibinfo {note} {(unpublished)}\BibitemShut {NoStop}%
\bibitem [{\citenamefont {Stefanucci}\ and\ \citenamefont {van
  Leeuwen}(2013)}]{stefanucci_van_leeuwen_2013}%
  \BibitemOpen
  \bibfield  {author} {\bibinfo {author} {\bibfnamefont {G.}~\bibnamefont
  {Stefanucci}}\ and\ \bibinfo {author} {\bibfnamefont {R.}~\bibnamefont {van
  Leeuwen}},\ }\href {\doibase 10.1017/CBO9781139023979} {\emph {\bibinfo
  {title} {Nonequilibrium Many-Body Theory of Quantum Systems: A Modern
  Introduction}}}\ (\bibinfo  {publisher} {Cambridge University Press},\
  \bibinfo {address} {Cambridge},\ \bibinfo {year} {2013})\BibitemShut
  {NoStop}%
\bibitem [{Note2()}]{Note2}%
  \BibitemOpen
  \bibinfo {note} {The Fock term of the self-energy disappears due to the Pauli
  exclusion principle.}\BibitemShut {Stop}%
\bibitem [{\citenamefont {Wei{\ss}e}\ \emph {et~al.}(2006)\citenamefont
  {Wei{\ss}e}, \citenamefont {Wellein}, \citenamefont {Alvermann},\ and\
  \citenamefont {Fehske}}]{Weisse2006RMP}%
  \BibitemOpen
  \bibfield  {author} {\bibinfo {author} {\bibfnamefont {A.}~\bibnamefont
  {Wei{\ss}e}}, \bibinfo {author} {\bibfnamefont {G.}~\bibnamefont {Wellein}},
  \bibinfo {author} {\bibfnamefont {A.}~\bibnamefont {Alvermann}}, \ and\
  \bibinfo {author} {\bibfnamefont {H.}~\bibnamefont {Fehske}},\ }\href
  {\doibase 10.1103/RevModPhys.78.275} {\bibfield  {journal} {\bibinfo
  {journal} {Rev. Mod. Phys.}\ }\textbf {\bibinfo {volume} {78}},\ \bibinfo
  {pages} {275} (\bibinfo {year} {2006})}\BibitemShut {NoStop}%
\bibitem [{\citenamefont {Holzner}\ \emph {et~al.}(2011)\citenamefont
  {Holzner}, \citenamefont {Weichselbaum}, \citenamefont {McCulloch},
  \citenamefont {Schollw\"ock},\ and\ \citenamefont {von
  Delft}}]{Holzner2011PRB}%
  \BibitemOpen
  \bibfield  {author} {\bibinfo {author} {\bibfnamefont {A.}~\bibnamefont
  {Holzner}}, \bibinfo {author} {\bibfnamefont {A.}~\bibnamefont
  {Weichselbaum}}, \bibinfo {author} {\bibfnamefont {I.~P.}\ \bibnamefont
  {McCulloch}}, \bibinfo {author} {\bibfnamefont {U.}~\bibnamefont
  {Schollw\"ock}}, \ and\ \bibinfo {author} {\bibfnamefont {J.}~\bibnamefont
  {von Delft}},\ }\href {\doibase 10.1103/PhysRevB.83.195115} {\bibfield
  {journal} {\bibinfo  {journal} {Phys. Rev. B}\ }\textbf {\bibinfo {volume}
  {83}},\ \bibinfo {pages} {195115} (\bibinfo {year} {2011})}\BibitemShut
  {NoStop}%
\bibitem [{\citenamefont {Jarrell}(1992)}]{Jarrell_1992}%
  \BibitemOpen
  \bibfield  {author} {\bibinfo {author} {\bibfnamefont {M.}~\bibnamefont
  {Jarrell}},\ }\href {\doibase 10.1103/PhysRevLett.69.168} {\bibfield
  {journal} {\bibinfo  {journal} {Phys. Rev. Lett.}\ }\textbf {\bibinfo
  {volume} {69}},\ \bibinfo {pages} {168} (\bibinfo {year} {1992})}\BibitemShut
  {NoStop}%
\bibitem [{\citenamefont {Hoshino}\ and\ \citenamefont
  {Werner}(2015)}]{Hoshino_2015}%
  \BibitemOpen
  \bibfield  {author} {\bibinfo {author} {\bibfnamefont {S.}~\bibnamefont
  {Hoshino}}\ and\ \bibinfo {author} {\bibfnamefont {P.}~\bibnamefont
  {Werner}},\ }\href {\doibase 10.1103/PhysRevLett.115.247001} {\bibfield
  {journal} {\bibinfo  {journal} {Phys. Rev. Lett.}\ }\textbf {\bibinfo
  {volume} {115}},\ \bibinfo {pages} {247001} (\bibinfo {year}
  {2015})}\BibitemShut {NoStop}%
\bibitem [{\citenamefont {Gull}\ \emph {et~al.}(2011)\citenamefont {Gull},
  \citenamefont {Millis}, \citenamefont {Lichtenstein}, \citenamefont
  {Rubtsov}, \citenamefont {Troyer},\ and\ \citenamefont {Werner}}]{Gull_2011}%
  \BibitemOpen
  \bibfield  {author} {\bibinfo {author} {\bibfnamefont {E.}~\bibnamefont
  {Gull}}, \bibinfo {author} {\bibfnamefont {A.~J.}\ \bibnamefont {Millis}},
  \bibinfo {author} {\bibfnamefont {A.~I.}\ \bibnamefont {Lichtenstein}},
  \bibinfo {author} {\bibfnamefont {A.~N.}\ \bibnamefont {Rubtsov}}, \bibinfo
  {author} {\bibfnamefont {M.}~\bibnamefont {Troyer}}, \ and\ \bibinfo {author}
  {\bibfnamefont {P.}~\bibnamefont {Werner}},\ }\href {\doibase
  10.1103/RevModPhys.83.349} {\bibfield  {journal} {\bibinfo  {journal} {Rev.
  Mod. Phys.}\ }\textbf {\bibinfo {volume} {83}},\ \bibinfo {pages} {349}
  (\bibinfo {year} {2011})}\BibitemShut {NoStop}%
\bibitem [{Note3()}]{Note3}%
  \BibitemOpen
  \bibinfo {note} {In the case of the double-ladder correction, even though the
  largest contribution may be shifted, we use the same $U^\protect \text
  {ren}$, to enable a meaningful comparison.}\BibitemShut {Stop}%
\bibitem [{\citenamefont {Kanamori}(1963)}]{10.1143/PTP.30.275}%
  \BibitemOpen
  \bibfield  {author} {\bibinfo {author} {\bibfnamefont {J.}~\bibnamefont
  {Kanamori}},\ }\href {\doibase 10.1143/PTP.30.275} {\bibfield  {journal}
  {\bibinfo  {journal} {Prog. Theor. Phys.}\ }\textbf {\bibinfo {volume}
  {30}},\ \bibinfo {pages} {275} (\bibinfo {year} {1963})}\BibitemShut
  {NoStop}%
\bibitem [{\citenamefont {Vidberg}\ and\ \citenamefont
  {Serene}(1977)}]{vidberg_analytical_continuation_1977}%
  \BibitemOpen
  \bibfield  {author} {\bibinfo {author} {\bibfnamefont {H.}~\bibnamefont
  {Vidberg}}\ and\ \bibinfo {author} {\bibfnamefont {J.}~\bibnamefont
  {Serene}},\ }\href@noop {} {\bibfield  {journal} {\bibinfo  {journal} {J. Low
  Temp. Phys.}\ }\textbf {\bibinfo {volume} {29}},\ \bibinfo {pages} {179}
  (\bibinfo {year} {1977})}\BibitemShut {NoStop}%
\bibitem [{\citenamefont {Nilsson}\ \emph {et~al.}(2017)\citenamefont
  {Nilsson}, \citenamefont {Boehnke}, \citenamefont {Werner},\ and\
  \citenamefont {Aryasetiawan}}]{Nilsson_2017}%
  \BibitemOpen
  \bibfield  {author} {\bibinfo {author} {\bibfnamefont {F.}~\bibnamefont
  {Nilsson}}, \bibinfo {author} {\bibfnamefont {L.}~\bibnamefont {Boehnke}},
  \bibinfo {author} {\bibfnamefont {P.}~\bibnamefont {Werner}}, \ and\ \bibinfo
  {author} {\bibfnamefont {F.}~\bibnamefont {Aryasetiawan}},\ }\href {\doibase
  10.1103/PhysRevMaterials.1.043803} {\bibfield  {journal} {\bibinfo  {journal}
  {Phys. Rev. Materials}\ }\textbf {\bibinfo {volume} {1}},\ \bibinfo {pages}
  {043803} (\bibinfo {year} {2017})}\BibitemShut {NoStop}%
\bibitem [{\citenamefont {Karrasch}\ \emph {et~al.}(2014)\citenamefont
  {Karrasch}, \citenamefont {Kennes},\ and\ \citenamefont
  {Moore}}]{PhysRevB.90.155104}%
  \BibitemOpen
  \bibfield  {author} {\bibinfo {author} {\bibfnamefont {C.}~\bibnamefont
  {Karrasch}}, \bibinfo {author} {\bibfnamefont {D.~M.}\ \bibnamefont
  {Kennes}}, \ and\ \bibinfo {author} {\bibfnamefont {J.~E.}\ \bibnamefont
  {Moore}},\ }\href {\doibase 10.1103/PhysRevB.90.155104} {\bibfield  {journal}
  {\bibinfo  {journal} {Phys. Rev. B}\ }\textbf {\bibinfo {volume} {90}},\
  \bibinfo {pages} {155104} (\bibinfo {year} {2014})}\BibitemShut {NoStop}%
\bibitem [{\citenamefont {Giamarchi}(2004)}]{giamarchi2004quantum}%
  \BibitemOpen
  \bibfield  {author} {\bibinfo {author} {\bibfnamefont {T.}~\bibnamefont
  {Giamarchi}},\ }\href@noop {} {\emph {\bibinfo {title} {{Q}uantum {P}hysics
  in {O}ne {D}imension}}}\ (\bibinfo  {publisher} {Oxford university press},\
  \bibinfo {address} {Oxford},\ \bibinfo {year} {2004})\ \bibinfo {note}
  {{N}ote: see Fig. 7.12(b) of this book. In our data, the gap is not apparent
  due to the small gap size and the peak broadening, but for much larger $U$,
  the spectrum of the optical conductivity is as depicted in this
  figure.}\BibitemShut {Stop}%
\bibitem [{\citenamefont {Shastry}\ and\ \citenamefont
  {Sutherland}(1990)}]{PhysRevLett.65.243}%
  \BibitemOpen
  \bibfield  {author} {\bibinfo {author} {\bibfnamefont {B.~S.}\ \bibnamefont
  {Shastry}}\ and\ \bibinfo {author} {\bibfnamefont {B.}~\bibnamefont
  {Sutherland}},\ }\href {\doibase 10.1103/PhysRevLett.65.243} {\bibfield
  {journal} {\bibinfo  {journal} {Phys. Rev. Lett.}\ }\textbf {\bibinfo
  {volume} {65}},\ \bibinfo {pages} {243} (\bibinfo {year} {1990})}\BibitemShut
  {NoStop}%
\bibitem [{\citenamefont {Rohringer}\ \emph {et~al.}(2018)\citenamefont
  {Rohringer}, \citenamefont {Hafermann}, \citenamefont {Toschi}, \citenamefont
  {Katanin}, \citenamefont {Antipov}, \citenamefont {Katsnelson}, \citenamefont
  {Lichtenstein}, \citenamefont {Rubtsov},\ and\ \citenamefont
  {Held}}]{RevModPhys.90.025003}%
  \BibitemOpen
  \bibfield  {author} {\bibinfo {author} {\bibfnamefont {G.}~\bibnamefont
  {Rohringer}}, \bibinfo {author} {\bibfnamefont {H.}~\bibnamefont
  {Hafermann}}, \bibinfo {author} {\bibfnamefont {A.}~\bibnamefont {Toschi}},
  \bibinfo {author} {\bibfnamefont {A.~A.}\ \bibnamefont {Katanin}}, \bibinfo
  {author} {\bibfnamefont {A.~E.}\ \bibnamefont {Antipov}}, \bibinfo {author}
  {\bibfnamefont {M.~I.}\ \bibnamefont {Katsnelson}}, \bibinfo {author}
  {\bibfnamefont {A.~I.}\ \bibnamefont {Lichtenstein}}, \bibinfo {author}
  {\bibfnamefont {A.~N.}\ \bibnamefont {Rubtsov}}, \ and\ \bibinfo {author}
  {\bibfnamefont {K.}~\bibnamefont {Held}},\ }\href {\doibase
  10.1103/RevModPhys.90.025003} {\bibfield  {journal} {\bibinfo  {journal}
  {Rev. Mod. Phys.}\ }\textbf {\bibinfo {volume} {90}},\ \bibinfo {pages}
  {025003} (\bibinfo {year} {2018})}\BibitemShut {NoStop}%
\bibitem [{\citenamefont {Gunnarsson}\ \emph {et~al.}(2017)\citenamefont
  {Gunnarsson}, \citenamefont {Rohringer}, \citenamefont {Sch\"afer},
  \citenamefont {Sangiovanni},\ and\ \citenamefont
  {Toschi}}]{PhysRevLett.119.056402}%
  \BibitemOpen
  \bibfield  {author} {\bibinfo {author} {\bibfnamefont {O.}~\bibnamefont
  {Gunnarsson}}, \bibinfo {author} {\bibfnamefont {G.}~\bibnamefont
  {Rohringer}}, \bibinfo {author} {\bibfnamefont {T.}~\bibnamefont
  {Sch\"afer}}, \bibinfo {author} {\bibfnamefont {G.}~\bibnamefont
  {Sangiovanni}}, \ and\ \bibinfo {author} {\bibfnamefont {A.}~\bibnamefont
  {Toschi}},\ }\href {\doibase 10.1103/PhysRevLett.119.056402} {\bibfield
  {journal} {\bibinfo  {journal} {Phys. Rev. Lett.}\ }\textbf {\bibinfo
  {volume} {119}},\ \bibinfo {pages} {056402} (\bibinfo {year}
  {2017})}\BibitemShut {NoStop}%
\bibitem [{\citenamefont {Chalupa}\ \emph {et~al.}(2018)\citenamefont
  {Chalupa}, \citenamefont {Gunacker}, \citenamefont {Sch\"afer}, \citenamefont
  {Held},\ and\ \citenamefont {Toschi}}]{PhysRevB.97.245136}%
  \BibitemOpen
  \bibfield  {author} {\bibinfo {author} {\bibfnamefont {P.}~\bibnamefont
  {Chalupa}}, \bibinfo {author} {\bibfnamefont {P.}~\bibnamefont {Gunacker}},
  \bibinfo {author} {\bibfnamefont {T.}~\bibnamefont {Sch\"afer}}, \bibinfo
  {author} {\bibfnamefont {K.}~\bibnamefont {Held}}, \ and\ \bibinfo {author}
  {\bibfnamefont {A.}~\bibnamefont {Toschi}},\ }\href {\doibase
  10.1103/PhysRevB.97.245136} {\bibfield  {journal} {\bibinfo  {journal} {Phys.
  Rev. B}\ }\textbf {\bibinfo {volume} {97}},\ \bibinfo {pages} {245136}
  (\bibinfo {year} {2018})}\BibitemShut {NoStop}%
\bibitem [{\citenamefont {Maki}(1968)}]{10.1143/PTP.40.193}%
  \BibitemOpen
  \bibfield  {author} {\bibinfo {author} {\bibfnamefont {K.}~\bibnamefont
  {Maki}},\ }\href {\doibase 10.1143/PTP.40.193} {\bibfield  {journal}
  {\bibinfo  {journal} {Prog. Theor. Phys.}\ }\textbf {\bibinfo {volume}
  {40}},\ \bibinfo {pages} {193} (\bibinfo {year} {1968})}\BibitemShut
  {NoStop}%
\bibitem [{\citenamefont {Thompson}(1970)}]{PhysRevB.1.327}%
  \BibitemOpen
  \bibfield  {author} {\bibinfo {author} {\bibfnamefont {R.~S.}\ \bibnamefont
  {Thompson}},\ }\href {\doibase 10.1103/PhysRevB.1.327} {\bibfield  {journal}
  {\bibinfo  {journal} {Phys. Rev. B}\ }\textbf {\bibinfo {volume} {1}},\
  \bibinfo {pages} {327} (\bibinfo {year} {1970})}\BibitemShut {NoStop}%
\bibitem [{\citenamefont {Aslamazov}\ and\ \citenamefont
  {Larkin}(1968)}]{Aslamazov1968EffectOF}%
  \BibitemOpen
  \bibfield  {author} {\bibinfo {author} {\bibfnamefont {L.~G.}\ \bibnamefont
  {Aslamazov}}\ and\ \bibinfo {author} {\bibfnamefont {A.}~\bibnamefont
  {Larkin}},\ }\href@noop {} {\bibfield  {journal} {\bibinfo  {journal} {{S}ov.
  {P}hys. {S}olid {S}tate}\ }\textbf {\bibinfo {volume} {10}},\ \bibinfo
  {pages} {875} (\bibinfo {year} {1968})}\BibitemShut {NoStop}%
\bibitem [{\citenamefont {Prokof'ev}\ and\ \citenamefont
  {Svistunov}(2007)}]{Prokofev_2004}%
  \BibitemOpen
  \bibfield  {author} {\bibinfo {author} {\bibfnamefont {N.}~\bibnamefont
  {Prokof'ev}}\ and\ \bibinfo {author} {\bibfnamefont {B.}~\bibnamefont
  {Svistunov}},\ }\href {\doibase 10.1103/PhysRevLett.99.250201} {\bibfield
  {journal} {\bibinfo  {journal} {Phys. Rev. Lett.}\ }\textbf {\bibinfo
  {volume} {99}},\ \bibinfo {pages} {250201} (\bibinfo {year}
  {2007})}\BibitemShut {NoStop}%
\bibitem [{\citenamefont {{Van Houcke}}\ \emph {et~al.}(2010)\citenamefont
  {{Van Houcke}}, \citenamefont {Kozik}, \citenamefont {Prokof'ev},\ and\
  \citenamefont {Svistunov}}]{Vanhoucke_2010}%
  \BibitemOpen
  \bibfield  {author} {\bibinfo {author} {\bibfnamefont {K.}~\bibnamefont {{Van
  Houcke}}}, \bibinfo {author} {\bibfnamefont {E.}~\bibnamefont {Kozik}},
  \bibinfo {author} {\bibfnamefont {N.}~\bibnamefont {Prokof'ev}}, \ and\
  \bibinfo {author} {\bibfnamefont {B.}~\bibnamefont {Svistunov}},\ }\href
  {\doibase https://doi.org/10.1016/j.phpro.2010.09.034} {\bibfield  {journal}
  {\bibinfo  {journal} {Phys. Procedia}\ }\textbf {\bibinfo {volume} {6}},\
  \bibinfo {pages} {95 } (\bibinfo {year} {2010})}\BibitemShut {NoStop}%
\end{thebibliography}%
\end{document}